\documentclass[]{aastex62}

\usepackage{graphicx}
\usepackage{float}
\usepackage[caption=false]{subfig}
\usepackage{enumitem}

\newcommand{\degree}{^\circ}
\newcommand{\settlingTime}{5-15 minutes}

\shorttitle{NIRCam Defocused Photometry}
\shortauthors{Schlawin et al.}

\begin{document}

\title{JWST NIRCam Defocused Imaging: Photometric Stability Performance and How it Can Sense Mirror Tilts}

\correspondingauthor{Everett Schlawin}
\email{eas342 AT EMAIL Dot Arizona .edu}

\author[0000-0001-8291-6490]{Everett Schlawin}
\affiliation{Steward Observatory \\
933 North Cherry Avenue \\
Tucson, AZ 85721, USA}

\author{Thomas Beatty}
\affiliation{Department of Astronomy\\
University of Wisconsin\\
Madison, Madison, WI 53706}
\author{Brian Brooks}
\affiliation{Space Telescope Science Institute\\
3700 San Martin Drive\\
Baltimore, MD 21218}
\author{Nikolay K. Nikolov}
\affiliation{Space Telescope Science Institute\\
3700 San Martin Drive\\
Baltimore, MD 21218}
\author[0000-0002-8963-8056]{Thomas P. Greene}
\affiliation{NASA Ames Research Center, Space Science and Astrobiology Division, MS 245-6, Moffett Field, CA, 94035, USA}
\author[0000-0001-9513-1449]{N\'estor Espinoza}
\affiliation{Space Telescope Science Institute\\
3700 San Martin Drive\\
Baltimore, MD 21218}
\affiliation{Department of Physics \& Astronomy, Johns Hopkins University, Baltimore, MD 2121818, USA}
\author[0000-0003-4669-7088]{Kayli Glidic}
\affiliation{Steward Observatory \\
933 North Cherry Avenue \\
Tucson, AZ 85721, USA}
\author{Keith Baka}
\affiliation{Anton Pannekoek Institute, University of Amsterdam, Science Park 904, 1098 XH Amsterdam, The Netherlands
}
\author{Eiichi Egami}
\affiliation{Steward Observatory \\
933 North Cherry Avenue \\
Tucson, AZ 85721, USA}
\author{John Stansberry}
\affiliation{Space Telescope Science Institute\\
3700 San Martin Drive\\
Baltimore, MD 21218}
\author{Martha Boyer}
\affiliation{Space Telescope Science Institute\\
3700 San Martin Drive\\
Baltimore, MD 21218}
\author{Mario Gennaro}
\affiliation{Space Telescope Science Institute\\
3700 San Martin Drive\\
Baltimore, MD 21218}
\affiliation{Department of Physics and Astronomy\\
Johns Hopkins University\\
3400 North Charles Street, Baltimore, MD 21218, USA}
\author[0000-0002-0834-6140]{Jarron Leisenring}
\affiliation{Steward Observatory \\
933 North Cherry Avenue \\
Tucson, AZ 85721, USA}
\author{Bryan Hilbert}
\affiliation{Space Telescope Science Institute\\
3700 San Martin Drive\\
Baltimore, MD 21218}
\author{Karl Misselt}
\affiliation{Steward Observatory \\
933 North Cherry Avenue \\
Tucson, AZ 85721, USA}
\author{Doug Kelly}
\affiliation{Steward Observatory \\
933 North Cherry Avenue \\
Tucson, AZ 85721, USA}
\author{Alicia Canipe}
\affiliation{Space Telescope Science Institute\\
3700 San Martin Drive\\
Baltimore, MD 21218}
\author[0000-0002-5627-5471]{Charles Beichman}
\affiliation{NASA Exoplanet Science Institute/IPAC\\
Jet Propulsion Laboratory, California  Institute of Technology\\
1200 E California Blvd\\
Pasadena, CA 91125}
\author{Matteo Correnti}
\affiliation{Space Telescope Science Institute\\
3700 San Martin Drive\\
Baltimore, MD 21218}

\author{J. Scott Knight}
\affiliation{Ball Aerospace \& technologies, Corp,
1600 Commerce Street,
Boulder, CO 80301}
\author[0000-0002-6880-4924]{Alden Jurling}
\affiliation{NASA Goddard Space Flight Center, 8800 Greenbelt Rd, Greenbelt, MD 20771, USA}
\author{Marshall D. Perrin}
\affiliation{Space Telescope Science Institute\\
3700 San Martin Drive\\
Baltimore, MD 21218}
\author{Lee D. Feinberg}
\affiliation{NASA Goddard Space Flight Center, 8800 Greenbelt Rd, Greenbelt, MD 20771, USA}
\author[0000-0003-0241-8956]{Michael W. McElwain}
\affiliation{NASA Goddard Space Flight Center, 8800 Greenbelt Rd, Greenbelt, MD 20771, USA}

\author{Nicholas Bond}
\affiliation{ADNET Systems, Inc./Code 550, NASA’s Goddard Space Flight Center, Greenbelt, MD 20771}

\author{David Ciardi}
\affiliation{NASA Exoplanet Science Institute/IPAC\\
Jet Propulsion Laboratory, California  Institute of Technology\\
1200 E California Blvd\\
Pasadena, CA 91125}

\author{Sarah Kendrew}
\affiliation{European Space Agency, Space Telescope Science Institute, 3700 San Martin Drive, Baltimore MD 21218}

\author[0000-0002-7893-6170]{Marcia Rieke}
\affiliation{Steward Observatory \\
933 North Cherry Avenue \\
Tucson, AZ 85721, USA}



\begin{abstract}

We use JWST NIRCam short wavelength photometry to capture a transit lightcurve of the exoplanet HAT-P-14~b to assess performance as part of instrument commissioning.
The short wavelength precision is 152 ppm per 27 second integration as measured over the full time series compared to a theoretical limit of 107 ppm, after corrections to spatially correlated 1/f noise.
Persistence effects from charge trapping are well fit by an exponential function with short characteristic timescales, settling on the order of \settlingTime.
The short wavelength defocused photometry is also uniquely well suited to measure the realtime wavefront error of JWST.
Analysis of the images and reconstructed wavefront maps indicate that two different hexagonal primary mirror segments exhibited ``tilt events" where they changed orientation rapidly in less than $\sim$1.4 seconds.
In some cases, the magnitude and timing of the flux jumps caused by tilt events can be accurately predicted with a telescope model.
These tilt events can be sensed by simultaneous longer-wavelength NIRCam grism spectral images alone in the form of changes to the point spread function, diagnosed from the FWHM.
They can also be sensed with the FGS instrument from difference images.
Tilt events possibly from sudden releases of stress in the backplane structure behind the mirrors were expected during the commissioning period because they were found in ground-based testing.
Tilt events have shown signs of decreasing in frequency but have not disappeared completely.
The detectors exhibit some minor (less than 1\%) deviations from linear behavior in the first few groups of each integration, potentially impacting absolute fluxes and transit depths on bright targets where only a handful of groups are possible.
Overall, the noise is within 50\% of the theoretical photon noise and read noise.
This bodes well for high precision measurements of transiting exoplanets and other time variable targets.

\end{abstract}




\section{Introduction} \label{sec:intro}

JWST will provide transformational new studies of exoplanet atmospheres with its unprecedented view in the infrared \citep{ahrer2022WASP39bERS,greene2016jwst_trans,morley2017temperateEarthSizedJWST,schlawin2018JWSTforecasts,bean2018ers}.
The NIRCam instrument's imaging, coronagraphic and spectrographic modes are providing powerful new insights into a range of astrophysical environments from the Solar System to the most distant galaxies from 0.6~\micron\ to 5.0~\micron\ \citep{rieke2005nircamSPIE}.
The NIRCam instrument has two wavelength channels that simultaneously observe the same field.
NIRCam's grism time series mode produces simultaneous spectroscopy on the long wavelength channel and photometry on the short wavelength channel.
The long wavelength spectroscopy employs a grism that enables spectroscopy from 2.4 to 5.0~\micron\ at R=1100 to 1700, with the wavelength coverage determined by the filter selected on the filter wheel \citep{greene2017jatisNIRCam}.
This wavelength range includes many prominent features in exoplanet atmospheres including H$_2$O, CH$_4$, CO, CO$_2$ and NH$_3$ \citep{greene2016jwst_trans}.
The molecular, atomic and ionic spectral features, when detected in exoplanet atmospheres, can give clues about the formation of planets, composition, dynamics, structure and cloud and haze composition \citep[e.g.][]{bitsch2021dryOrWaterSubNeptune,mordasini2016planetFormationSpec,powell2019transitSignaturesHotJups,kempton2017hazeVsCloud}.

Given the small scale heights of planetary atmospheres ($\lesssim 10^{-3} R_\odot$) compared to their host star radii ($\sim1 R_\odot$) and their low temperatures compared to their host star temperatures ($T_p^4 /T_*^4 \lesssim 10^{-2}$, where $T_p$ and $T_*$ are the temperatures of the planet and star respectively), exoplanet transit spectroscopy requires extreme precision.
The largest signals from even the most favorable planets are still very small. The `prominent' 1.4 $\mu$m water vapor feature strength varies from 100 ppm in the prototypical hot Jupiter HD 209458 b \citep{deming13} down to less than 30 ppm for the super-Earth GJ 1214 b \citep{kreidberg2018wasp103}.
The most favorable CO$_2$ signatures are predicted to be 50 ppm and below in the temperate Earth-sized TRAPPIST-1 d and e \citep{barstow2016trappist1habitable,lustig-yaeger2019detectabilityTRAPPIST-1}.
Lessons from previous space telescopes such as Spitzer, Kepler and the Hubble Space Telescope (HST) have taught us that many instrument systematics can be present including intrapixel sensitivity, thermal breathing, charge trapping, and pointing errors \citep{beichman2014pasp}.

Here we present the performance of the NIRCam short-wavelength time-series photometry based on commissioning observations of the HAT-P-14 system. A companion paper, \citet{beatty2022hatp14Spec} describes the simultaneous long-wavelength grism spectroscopy.
The light that enters NIRCam is split by a dichroic beamsplitter so that the short wavelengths (0.6 to 2.3~\micron) and long wavelengths (2.4 to 5.0~\micron) are directed to different channels' optics and detectors.
The pupil wheel and filter wheel selections determine the wavelength extents of the observations ( a short wavelength F210M filter paired with the long wavelength filter F322W2 will cover approximately 1.99 to 2.23~\micron\ on the short wavelength channel and 2.43 to 4.01~\micron\ on the long wavelength channel).
The flight software is configured to run the detectors with the same size subarrays and readout patterns so the two exposures are commanded to be simultaneous to within one frame time.
For the grism time series mode, two short wavelength detectors (NRCA1 and NRCA3) centered on nearly the same piece of sky as the long wavelength (NRCALONG) are read out by detector electronics and downloaded from the observatory. 
Appendix \ref{sec:subarrayPos} shows the relative layouts of the detectors.\footnote{Also see \url{https://jwst-docs.stsci.edu/jwst-near-infrared-camera/nircam-instrumentation/nircam-detector-overview} for the detector layout \citep{jdox2016}.}

When performing grism spectroscopy, the short wavelength channel can be (currently) configured to use the Weak Lens +4 or Weak Lens +8 lens optics, which are designed for mirror wavefront sensing \citep{greene2010jwstNIRCam}.
The Weak Lens +4 and Weak Lens +8 lens optics are a subset of all defocusing lenses, which also include Weak Lens -4 and Weak Lens -8.
The Weak Lens +4 (on the filter wheel) and Weak Lens +/- 8 (on the pupil wheel) can be used in series simultaneously so that focuses of -8, -4, 0, +4 and +12 Peak-to-Valley (P-V) waves can be configured \citep{greene2010jwstNIRCam}.
The lenses were tuned to be sensitive to various spatial frequencies and levels of aberration in the wavefront at 2.12~\micron\ \citep{dean2003diversitySelectionForPhaseRetrieval,perrin2016preparingJWSTwavefrontSensing}.
The Weak Lens +4 defocuses the light by 4.0 P-V units at 2.12~\micron\ and results in a core PSF of approximately 31 pixels and the Weak Lens +8 defocuses the light by 8.0 P-V units at 2.12~\micron\ and results in a core PSF of approximately 65 pixels.
Physically, the Weak Lens +8 defocuses the light by modifying the wavefront by +8.0 waves P-V at 2.12 um at the exit pupil, where the positive convention is that the phase at the edge of the wavefront leads the phase at the center of the pupil.
We assume that the equation for the phase in waves for the Weak Lens +8 is $8*( r^2 -0.5)$ where r is the exit pupil radius normalized to unity at the edge and 0 in the center.
This wavefront error displaces the best focus along the chief ray.

The defocused PSF from a weak lens makes the short wavelength imaging very sensitive to changes in the wavefront and it also spreads the light over more pixels to decrease the rate of saturation for a given star brightness.
While not yet supported, there is also an active proposal to allow the use of the Dispersed Hartmann Sensor (DHS) in the short wavelength channel instead of the weak lenses.
Utilizing the DHS in the short wavelength channel would allow simultaneous spectroscopy across a large fraction of the full NIRCam wavelength regime \citep{schlawin2017dhs}.

The target HAT-P-14 b \citep{torres2010hatp14} was selected for calibration purposes because it has no strongly detectable atmospheric features and a flat spectrum due to a high gravity \citep[42 m s$^{-2}$, calculated from ][]{stassun2017gaiaRadiiMasses} and small atmospheric scale height (2$\times10^{-4}$ R$_\odot$).
The 3.4 M$_{J}$ 1.4 R$_{J}$ 1600 K planet \citep{stassun2017gaiaRadiiMasses,southworth2012homogenousStudies38Planets} orbits a $K=8.85$, 6600 K F-type star with low chromospheric activity \citep{torres2010hatp14}.
It was observed by JWST NIRSpec \citep{espinoza2022hatp14Spec}, NIRISS and NIRCam to enable cross-instrument performance checks and to determine the transit depths are consistent.
The expected variations in the transmission spectrum are 16 ppm, assuming a typical atmosphere that exhibits atmospheric features that are equivalent to 0.9 atmospheric scale heights \citep{wakeford2019atmosphericForecastMuted}, so the differences across wavelengths are expected to be about as large as the theoretical ideal photon noise $\sim$9 ppm for these observations.
One option to characterize stability is to observe a known quiescent star and verify that the lightcurves are flat with time.
However, in the event that the lightcurve is not perfectly flat, disentangling the measured precision from the baseline trend curves is challenging.
Any de-trending done to the data, such as a Gaussian process regression, can remove trends but could also remove astrophysical signal on a time variable source.
Therefore, we observed a known transiting planet with a known ephemeris to determine the ability of JWST to measure the transit depth.
While the target was chosen to have no significant atmospheric signal (no significant change in transit depth with wavelength), it still has a transit depth of 6480 $\pm$ 240 ppm \citep{torres2010hatp14} or about 0.6\% which is readily detectable.

\begin{figure*}
\gridline{\fig{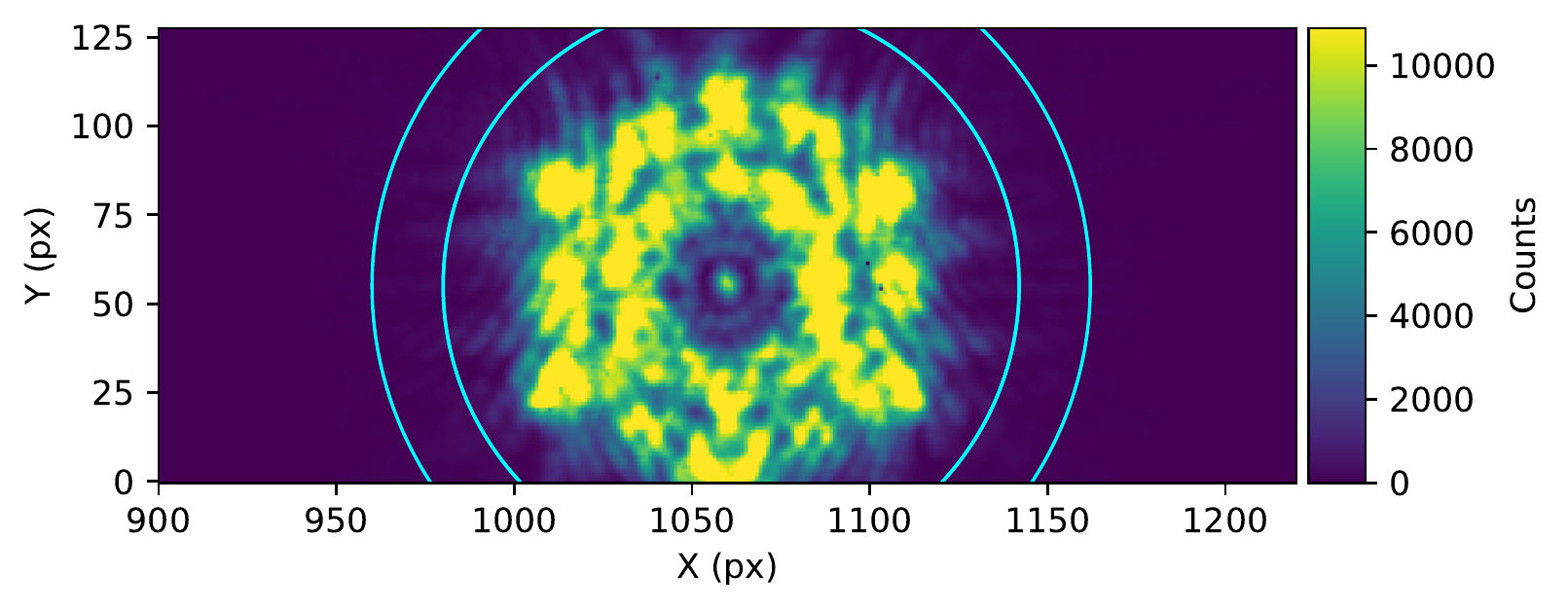}{0.68\textwidth}{Weak Lens +8 PSF}}
\caption{The Weak Lens +8 PSF spreads the light over a large region, which reduces the chance of saturation and reduces intra-pixel sensitivity.
The Weak Lens +8 was designed for retrieving the primary mirror state and thus phase retrieval with the Weak Lens enables real-time monitoring of the mirror wavefront error.
The above image is near the middle of a 2048$\times$128 SUBGRISM128 subarray with the F210M filter,  a 27.0 second long integration time on HAT-P-14 with a plate scale of about 31 mas /px.
The background subtraction annulus as shown as concentric circles and is the inner radius is the same as the source aperture used for photometry.
The subarray position was adjusted downward after HAT-P-14 b commissioning to better center the hexagonal PSF on the subarray.\label{fig:wlp8PSF}}
\end{figure*}

Section \ref{sec:obs} describes the observations during commissioning and how they were analyzed.
Section \ref{sec:systematics} describes systematic effects in the data including mirror tilt events, tilt timescales, the charge trapping ramp and observed non-linearity.
Section \ref{sec:NoiseAndStabilityPerformance} describes the noise performance, pointing performance and time accuracy as measured from in-flight data and the transit depth precision when accounting for contamination from a background source.
We present our conclusions in Section \ref{sec:conclusions}.

\begin{deluxetable*}{ccc}[b!]
\tablecaption{Summary of Key Times During Observation\label{tab:observationTimes}}
\tablecolumns{2}
\tablewidth{0pt}
\tablehead{
\colhead{UT start time} &
\colhead{Integration Number} &
\colhead{Description} \\
\colhead{(YYYY-mm-ddThh:mm:ss)} & \colhead{}  & \colhead{} \\
}
\startdata
2022-05-02T06:43:04 & 1 & Exposure Start \\
2022-05-02T09:21:34 & 344 & Transit Start (Contact 1) \\
2022-05-02T10:10:01 & 449 & HGA Move \\
2022-05-02T10:30:20 & 492  & Transit Midpoint\\
2022-05-02T10:30:41 & 493 & Tilt Event 1 \\
2022-05-02T11:23:19 & 607 & Tilt Event 2 \\
2022-05-02T11:39:05 & 641 & Transit End (Contact 4) \\
2022-05-02T12:43:25 & 780 & Exposure End\\
\enddata
\tablecomments{Times are in Coordinated Universal Time. The barycentric dynamical times are approximately 4.63 minutes later.
Integration numbers are given for 1-based counting}
\end{deluxetable*}

\section{Observations and Image Processing}\label{sec:obs}

\subsection{Observation Description}\label{sec:obsDescrip}

We observed the hot Jupiter HAT-P-14 b to test the instrument performance and verify that the NIRCam Grism Time Series mode was suitable for science.
The planet was selected to have a short duration transit (2.3 hr), high brightness that enables high precision, low stellar activity and high planet surface gravity to have a small atmospheric signal.
The short duration requirement was selected to minimize the needed observing time but has the flip side of preferring targets with near-grazing transits (high impact parameters).
HAT-P-14 b's impact parameter is 0.91 \citep{fukui2016hatp14}, so fitting the limb darkening without the planet crossing through the stellar mid-point can complicate the transit depth measurement and stellar models may be preferred for the limb darkening law, as described in Section \ref{sec:lcFitting}.
The observation exposure started at 2022-05-02T06:43 UTC, which was the earliest start time allowed by the Astronomy Proposal Tool (APT) special requirements to ensure a long baseline before transit ingress.
The lightcurve was observation 1 of Commissioning Program ID 1442, which is available on the Barbara A. Mikulski Archive for Space Telescopes (MAST).
Table \ref{tab:observationTimes} lists key times during the observations.

The exposure duration was 6 hours, which enabled 2.6 hours of stabilization time before planet ingress.
The SW channel was configured with the Weak Lens+8 pupil, F210M filter which covers approximately 1.9 to 2.2 $\mu$m.
This was paired with the GRISMR pupil and F322W2 filter on the long wavelength channel, discussed in \citet{beatty2022hatp14Spec}.
The exposure setup was a single exposure with the BRIGHT2 readout pattern, 20 groups integrated up the ramp and 780 integrations for a 2048$\times$128 subarray (SUBGRISM128) with 4 output amplifier channels.
NIRCam readout electronics sample non-destructive reads (ie. frames) to create a ramp and then reset the detector at the end of an integration.
Detector groups are a set of 1 or more detector image frames where the frames have been either averaged or skipped by onboard electronics and the average frame (ie. a group) is downloaded from the spacecraft.
The averaging or else skipping of frames between groups is done to reduce the data volume on some observations.
Observations when more than one frame is averaged or skipped per group reduce the data volume as compared to saving every frame per integration.
In BRIGHT2, there are 2 frames (ie. detector samples) averaged per group and zero frames skipped.
This resulted in a frame time of 0.67597 seconds, a group time of 1.35194 seconds and a cadence of 27.72001 seconds per integration for 780 integrations.

The 6 hour observation was continuous with no interruptions.
Pointing performance was extremely stable as discussed in Section \ref{sec:pointingPerformance} below.
There was one high gain antenna (HGA) move during the exposure but no interruptions or pauses to the integrations in the 6 hour exposure.
This allows detector charge trapping effects to approach a steady state, unlike with HST where each satellite orbit around Earth results in a new charge trapping ramp effect approximately every 90 minutes \citep[e.g.][]{berta2012flat_gj1214}.

The point spread function (PSF) with the Weak Lens+8 pupil is shown in Figure \ref{fig:wlp8PSF}.
The Weak Lens +8 (along with the -8 wave and +4 wave lenses) is designed to measure the optical path differences of the primary mirror surface.
It was enabled as an option in the grism time series and imaging time series modes to spread out the light over more pixels and also reduce the saturation on bright targets.
Having the light spread out over more pixels reduces the intrapixel sensitivity, which strongly affected \textit{Spitzer Space Telescope} lightcurves because the PSF was highly undersampled \citep{ingalls2016spitzerRepeatability}.
It also allows realtime monitoring of the mirror surface along with the long wavelength grism spectroscopy, discussed in Section \ref{sec:tiltEvents}.
We note that the full extent of the PSF is asymmetrically truncated by the subarray.
This will be mitigated in Cycle 1 science observations because an update was made to better center the subarray position.
Even with better centering, the SUBGRISM128 subarray truncates some of the wings of the PSF.
In other words, the commissioning observations (Figure \ref{fig:wlp8PSF}) had the PSF is asymmetrically truncated and in science observations it will be still truncated in the wings but more symmetrically.
Fortunately, JWST pointing is so stable (to be discussed in Section \ref{sec:pointingPerformance}) that there were no impacts from the PSF truncation on the lightcurve and zero-padding the image for wavefront analysis resulted in high quality wavefront phase retrievals (to be discussed in Section \ref{sec:tiltEvents}).

\begin{figure*}
\gridline{\fig{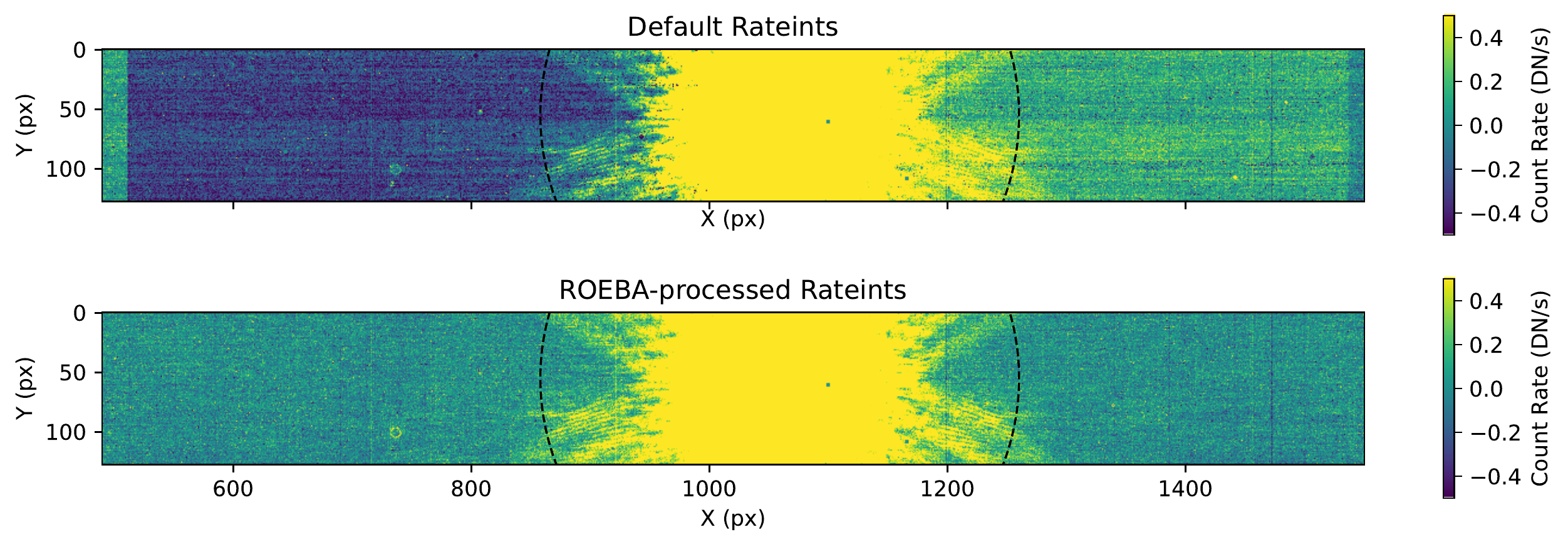}{0.99\textwidth}{}}
\caption{The default rateints product for integration number 83 with no extra corrections (top) has large pre-amplifier offsets (rectangular-shaped offsets) and 1/f noise (horizontal banding).
A row-by-row, odd/even by amplifier (ROEBA) correction, described in Section \ref{sec:pipeAdjustments} uses background pixels to reduce the pre-amplifier offset and 1/f noise.
The background pixels are the ones outside of the dashed black circle that has a radius of 201 pixels. \label{fig:roebaCor}}
\end{figure*}

\subsection{Data Analysis from Uncalibrated Data to Rates Per integration}\label{sec:pipeAdjustments}
We begin with the \texttt{uncal} data (also known as ``MAST level 1 uncalibrated'', ``User Data Product Stage 0 Fully-Populated FITS file'' or Data Processing Level 1b).
We then run Stage 1 of the Calwebb STScI pipeline \citep{bushouse2022jwstPipeline} with some modifications described below, to produce a \texttt{rateints} file which contains the counts in DN/s for each integration.

We started with the STScI JWST Detector 1 Stage 1 pipeline version 1.6.0 and the Calibration Reference Data System (CRDS) context \texttt{jwst\_0822.pmap} to convert the raw data values to signal slopes (DN/s) per integration.
The Detector Stage 1 pipeline begins by scaling the groups (which doesn't affect NIRCam readout modes), initializing the data quality flags, flagging saturated pixels and then subtracting a superbias.
When examining the superbias subtraction, we note that every 4 integrations has a bias offset structure due to the fast frame resets that reset pixels across the whole detector to ensure that charge doesn't migrate into the subarray.
For this SUBGRISM128 subarray, the fast frame reset occurs every 4 integrations.
However, accounting for the different superbias every 4 integrations resulted in very little change to the lightcurve so we used a single superbias subtraction for all integrations.

Normally, detector stage 1 processing then does a reference pixel correction step.
The reference pixels are light-insensitive pixels that share the same electronic noise sources as the regular light-sensitive pixels and thus allow efficient subtraction of common-mode noise sources.
Normally, the bottom and side reference pixels are used to mitigate odd/even column effects, pre-amplifier resets and 1/f noise.
However, the SUBGRISM128 subarray for the short-wavelength time series mode has no reference pixels along the top or bottom detector rows.\footnote{The SW subarray is designed to be centered on the defocused PSF when the same source is centered near the Y=34 long wavelength pixel on the LW subarray. See section \ref{sec:subarrayPos} for a diagram. The relative positions of the short and long wavelength detectors when projected on the sky result in the weak lens +8 image being centered at Y= 167.5 in 1-based counting absolute coordinates on the short wavelength detector. The subarray position was adjusted to better center the PSF after HAT-P-14 was observed.}
Thus, the reference pixel step does not correct for large pre-amplifier offsets and smaller odd/even offsets.
We instead replace the reference pixel step with a similar step that uses background instead of reference pixels to remove many of the same effects.

We use a shorthand for this modified reference pixel step ROEBA, which stands for row-by-row, odd/even by amplifier correction, which lists the steps in reverse order.
First, we select one of the four amplifiers for correction at a time.
In this case a block pixels from 1 to 512, 513 to 1024, 1025 to 1536 or 1537 to 2048 pixels in the X direction and the entire 128 pixel tall subarray in the Y direction.
For this correction, it is necessary to define a background region to use as a proxy for reference pixels that has minimal contamination from bright sources.
We select all pixels that are more than 201 pixels away from the central PSF as shown in Figure \ref{fig:roebaCor}.
We then apply a slow-read correction, which subtracts the median odd count level in the background region from all odd numbered columns and the median even count level from in the background region from all even numbered columns.
This removes the pre-amplifier offsets.
We then calculate the row-by-row median count level from the background region and subtract this from all pixels in that row.
This largely removes the longer time scale (low frequency) components of the 1/f noise \citep[e.g.][]{schlawin2020jwstNoiseFloorI}.
The resulting correction from ROEBA shown in Figure \ref{fig:roebaCor} (bottom) has substantially smaller pre-amplifier reset offsets in each amplifier as well as less 1/f noise banding along the horizontal direction.
We note that the ROEBA method uses many more pixels across the horizontal fast-read direction than the side reference pixels on the detector so it can improve results even when reference pixels are available in a subarray.
Another advantage of the ROEBA correction is that many fewer pixels are erroneously marked as jumps at the jump detection step (described below).
One assumption in the ROEBA step is that it relies on a clean background region and could potentially over-subtract when there is faint scattered light or background sources so inspection for neighboring sources and faint emission is necessary before applying it to all NIRCam data.

We next apply the non-linearity correction with default parameters.
The result of the bias subtraction and ROEBA step is that the signal level is close to 0 DN at the background before applying the linearity correction step.
The stage 1 pipeline normally has a dark current subtraction step after non-linearity correction, but dark current on the NIRCam short wavelength detectors is less than 0.05 e$^{-}$/sec and there is currently too much signal from cosmic rays and persistence to measure well and subtract on these detectors.
We next run the jump step correction with a threshold of 6 sigma.
Without the ROEBA correction, many pixels are erroneously flagged as ``jumps'' (e.g. cosmic rays) because the pre-amplifier offsets and 1/f noise can appear as jumps in signal from one detector group to another.
With the ROEBA correction, we note that these erroneous jumps are much less frequent and instead the jump step flags true outliers such as cosmic rays.
One exception where the jump step does not flag a cosmic ray perfectly is the outer radii of `snowball`` events \citep{rigby2022jwstPerformance}, as shown in Figure \ref{fig:roebaCor} near X=737,Y=101.
Here, only the brighter portions of the snowball are flagged with the jump detection algorithm used in the analysis for this paper.
As the final step of stage 1, we run the ramp fit step with default parameters to calculate a count rate per pixel in Data Numbers (DN) per second.
There is another gain scaling step for NIRSpec data but it does not affect the NIRCam data presented in this paper.

\subsection{Photometric Extraction}\label{sec:photExtract}

We next use aperture photometry with \texttt{photutils} \citep{bradley2016photutilsv0p3} to measure the flux for each integration to derive a flux as a function of time.
We fit the central peak of the PSF with a Gaussian and used this to determine the aperture center for each integration.
We use a circular source aperture with a radius of 81 pixels, and the mean background rate from a background annulus with an inner radius of 81 and an outer radius of 101 pixels.
These pixels were found to produce the minimum lightcurve scatter from an aperture size grid search.
Although the ROEBA step described above largely removes the background, we found that the additional background annulus subtraction improved the precision, likely because of residual 1/f not removed by ROEBA.
We found the same level of noise whether we did aperture centering or kept the apertures fixed, with the exception of better photometry during the HGA move affecting one integration to compensate for the small amount of image motion within that one integration.
We normalize the flux time series to a value near 1.0 (1000 parts per thousand, ppt) for lightcurve fitting and noise analysis, which we describe in Section \ref{sec:lcFitting}.

\section{Lightcurve Systematics}\label{sec:systematics}
The lightcurves overall exhibited very low level systematic errors, which we describe in this section.
There were changes to the PSF due to ``tilt'' events as discussed in Section \ref{sec:tiltEvents} through \ref{sec:tiltTimescale}, an exponential ramp discussed in Section \ref{sec:rampTimescale} and some evidence for non-linearity discussed in Section \ref{sec:nonLinearEffects}.

\subsection{Tilt Events}\label{sec:tiltEvents}

\begin{figure*}
\gridline{\fig{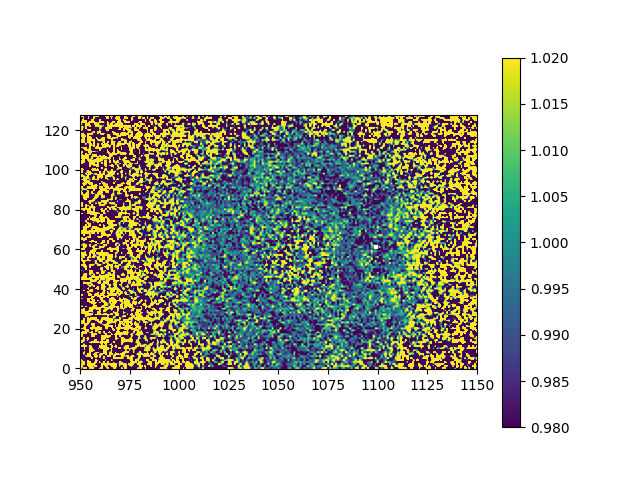}{0.33\textwidth}{Weak Lens Before Tilt 1}
          \fig{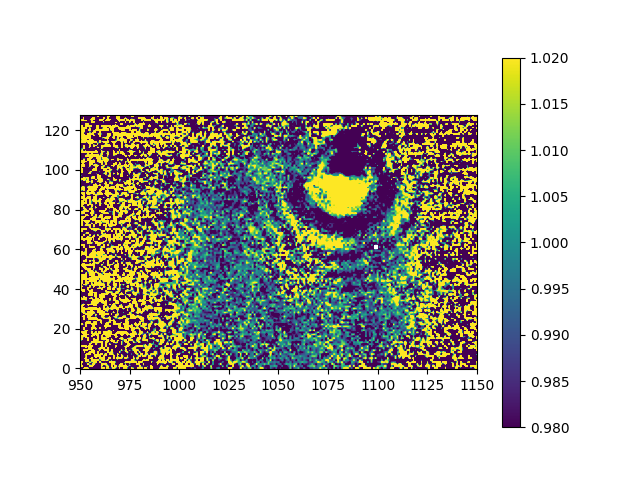}{0.33\textwidth}{Weak Lens After Tilt 1}
          \fig{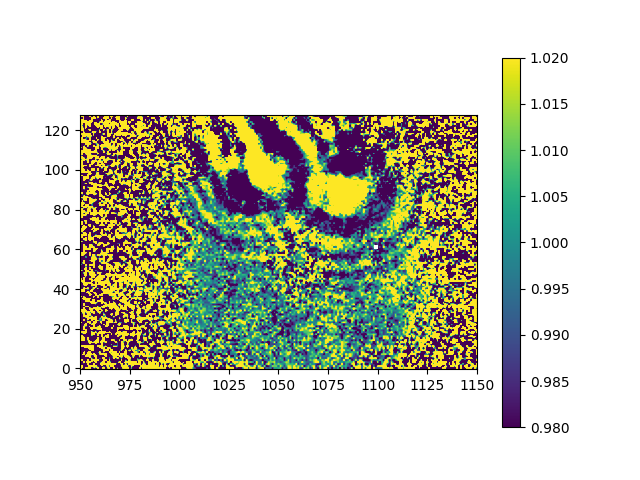}{0.33\textwidth}{Weak Lens After Tilt 2}
          }
     \gridline{\fig{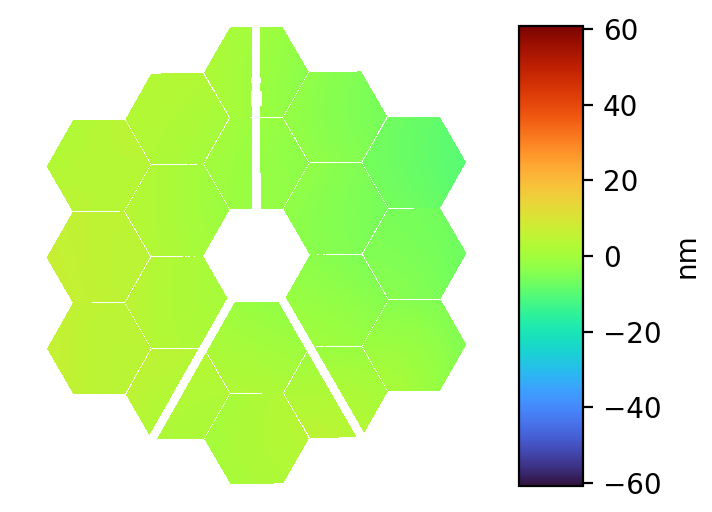}{0.23\textwidth}{Optical Path Difference Before Tilt 1}
          \fig{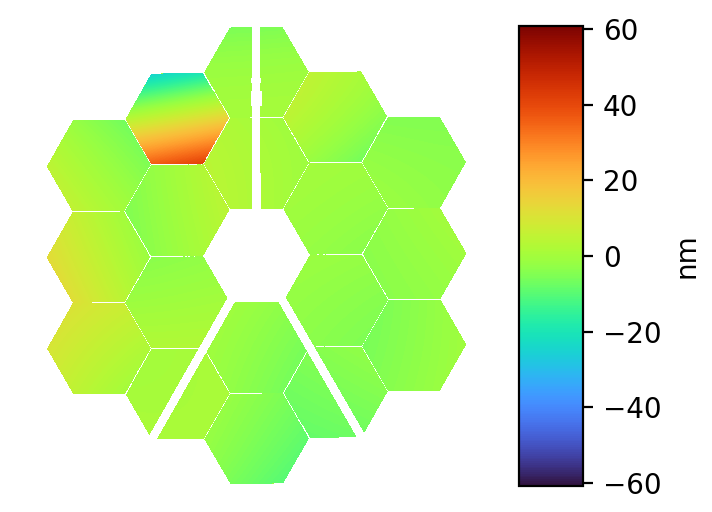}{0.23\textwidth}{Optical Path Difference After Tilt 1 - C6 tilt}
          \fig{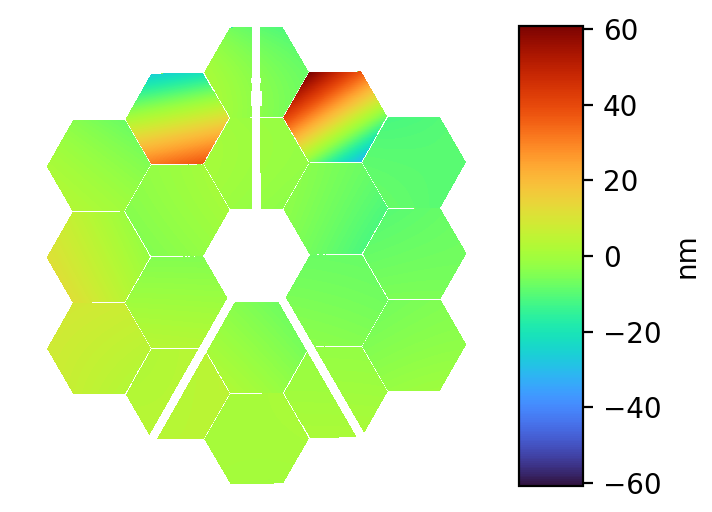}{0.23\textwidth}{Optical Path Difference After Tilt 2 - C1 tilt}
          }
\caption{Weak Lens Ratio Images before and after both Tilt Events (top) and Phase retrieval, oriented as if looking down at the mirror surface from the secondary mirror (bottom plots). \label{fig:tiltDiffImages}}
\end{figure*}

We originally noticed a large jump in the long wavelength time series \citep{beatty2022hatp14Spec} but also found one at the same time at some aperture sizes in the short wavelength and a smaller jump later in the time series.
The jumps in flux correspond to changes in the mirror surface, known as ``tilt events'' \citep{rigby2022jwstPerformance}.
These were first noticed by choosing an arbitrary reference image (in this case the rate from the 6th integration) and dividing the rate from all other integrations by it.
Figure \ref{fig:tiltDiffImages} shows three example ratio images from 1) before the tilt events, 2) after the first tilt event and 3) after the second tilt event.
The tilt events show up as clear wave patterns in these differential images due to changes in the mirror surface.
The short wavelength Weak Lens +8 defocused PSF is part of the wavefront sensing tools aboard JWST and is thus highly sensitive to changes in the mirror surface.

\begin{figure*}
\gridline{\fig{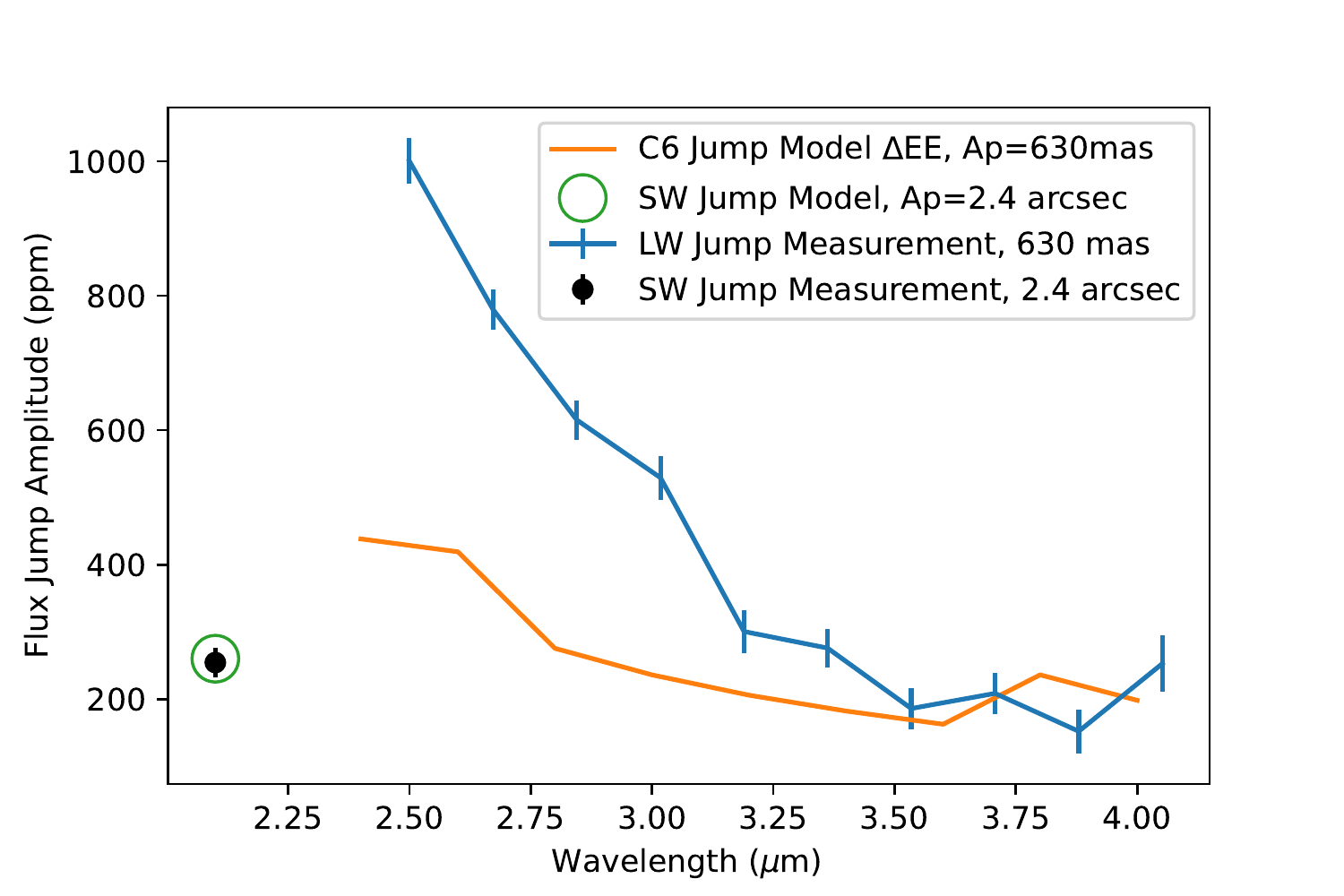}{0.85\textwidth}{Jump Spectrum and Models for the C6 Tilt Event}}
\caption{The jump in flux caused by the C6 tilt event is wavelength dependent.
The black point and blue points with error bars show the measured flux jump size for the short wavelength (SW) aperture of 2.4\arcsec\ and LW aperture of 0.630\arcsec, respectively.
Our current model (green circle) correctly predicts the SW channel's jump with a 2.4\arcsec\ aperture but under-predicts the LW 0.63\arcsec\ jump size (orange curve) for the grism spectrum, especially at 2.5 to 3.0~\micron.}\label{fig:jumpSpectrum}
\end{figure*}

We used a non-linear optimization based phase retrieval algorithm \citep{fienup1982, fienup1993a} based on the algorithmic differentiation \citep{jurling2014b}. The coherent propagation model used a flexibly sampled discrete Fourier transform \citep{jurling2018}. We used a bias and gain invariant error metric \citep{thurman2009a} as the cost function. The phase retrieval model was initialized using the wavefront and pupil data calculated from the JWST optical model for the nearest available  field point, the NRCA3-FP2MIMF multi-instrument multi-field point. Because PSF data were only available with the +8 wave weak lens, this was a single image phase retrieval problem. We  configured the model with 128 pixels across the array storing the pupil and 256 pixels across the cropped and padded image. Because of the band-pass of the F210M filter used for imaging, we used a broadband point-spread function model \citep{fienup1999, jurling2018}. We ran the phase retrieval optimization in two steps.

In the first step we used the first image in the time series to establish a starting point. We sequentially estimated image position (using a sub-pixel cross correlation method \citep{guizar-sicairos2008}), global low order Zernike aberrations, segment level piston, and tip/tilt errors (1st order segment Zernike aberrations). Finally, we performed joint estimation over a smoothly interpolated low resolution grid in wavefront and amplitude. With a single image the ability to distinguish between phase and amplitude errors in this last step is limited.

In the second step, we begin from the reference point established by the first phase retrieval. In this case the optimization was done jointly over global and segment Zernike aberrations without sequential bootstrapping. This facilitates change detection by requiring the pupil amplitude and higher order wavefront variation to be the same throughout the time series, but still allowing low order and segment wavefront changes through time. 

The phase retrieval results are shown in Figure \ref{fig:tiltDiffImages} and it is clear that the C6 mirror segment changed orientation (tilted) at event 1 and that the C1 mirror segment tilted at event 2.
In event 1, the wavefront over C6 changed by 18 nm RMS (evaluated over the segment) compared to the initial wavefront, and in event 2, C1 changed by 26 nm (again evaluated over the segment). These changes are much smaller than the wavelength of the light, so both the individual segments and the overall telescope remain diffraction limited.

As shown in Figure \ref{fig:jumpSpectrum}, the C6 event produced a larger change with a magnitude that varies from 200 ppm to 1000 ppm across the 2.4~\micron\ to 4.0~\micron\ wavelength range for the long wavelength and 254 $\pm$ 22 ppm in the short wavelength photometry using an aperture radius of 79 pixels or 2.45 arcsec.
Using the wavefront phase retrieval described in this section above \citep{jurling2014b} for the optical path differences, we make a prediction for the magnitude of the flux jump by comparing the enboxed or encircled energy before and after the jumps.
We use enboxed energy for the long wavelength spectroscopic extraction box and encircled energy for the short wavelength photometry.
The short wavelength (SW) model prediction of 260 ppm is very close to the measured 254 $\pm$ 22 ppm for the large aperture used in the short wavelength time series.
The long wavelength (LW) jump model prediction under-predicts the jump size from 2.5 to 3.5 $\mu$m but agrees from 2.7 to 4.1 $\mu$m.
Some of the difference between the model and measurement could be the difference between an imaging and dispersed PSF and how they are extracted.
Preliminary models with a dispersed LW grism image can reproduce the 1000 ppm change at 2.5 $\mu$m.

\subsection{Ways to Detect Tilt Events}\label{sec:tiltDetectionMethods}
\begin{figure*}
\gridline{\fig{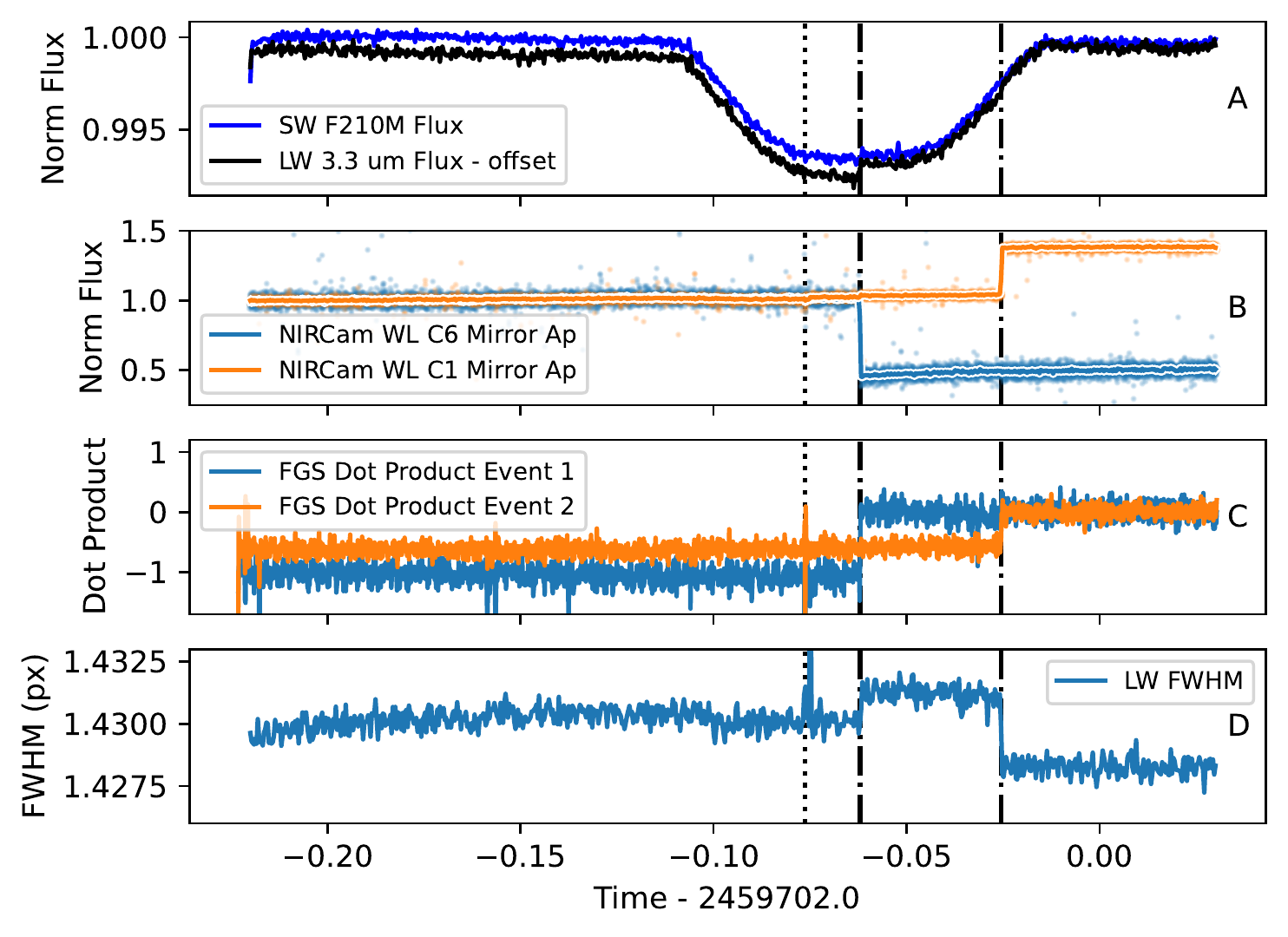}{0.6\textwidth}{Normalized Flux and Tilt-Sensitive Metrics}
\fig{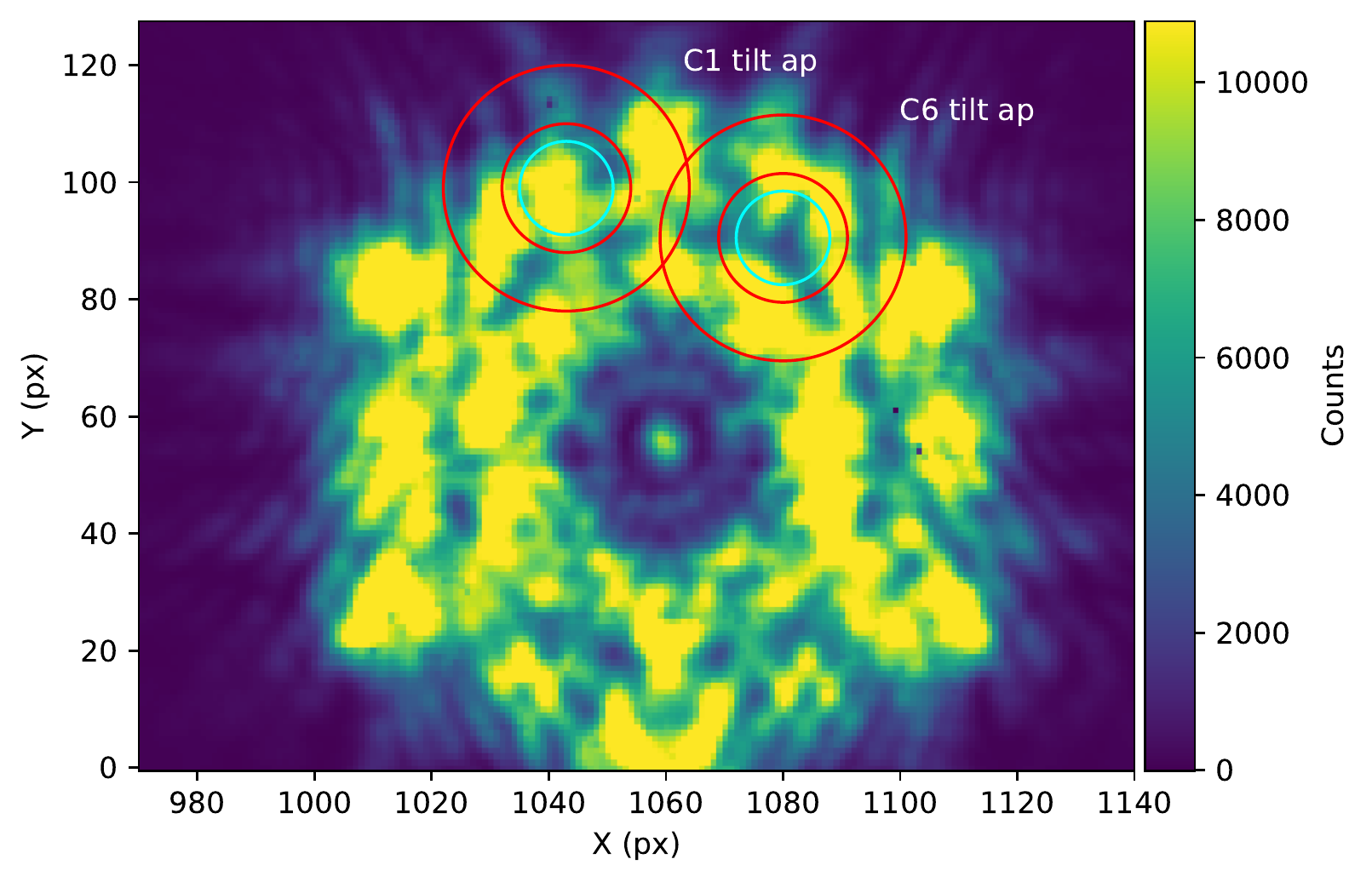}{0.3\textwidth}{Photometric Apertures on the C6 and C1 mirror segments}}
\caption{The lightcurves exhibited jumps most noticeably in the long wavelength spectroscopy, shown here as a broadband time series (plot A).
These jumps (marked as dash-dot vertical lines) can be detected in a variety of ways.
Apertures placed on the short wavelength weak lens (WL) PSF are highly sensitive to mirror tilts on mirrors C6 and C1 (plot B).
The time series are shown on a group-by-group difference image (points in blue and orange, plot B) as well as on the full jump-corrected integration (solid blue and orange lines, plot B).
The FGS instrument can also sense mirror tilts by a dot product of a difference images, as defined in Equation \ref{eq:diffDotProduct} (plot C).
The FWHM of the long wavelength spectroscopy can also be used to sense mirror tilts (plot D).
The HGA move (vertical dotted line) also can create spikes in some tilt statistics.
The apertures used to sense the C6 and C1 mirror1 tilts (right plot) are the ones used to generate the time series in plot B. \label{fig:mirrorPhotTser}}
\end{figure*}

It is possible to quantitatively measure tilt events in addition to visually inspecting the differential images or doing the phase retrieval shown in Figure \ref{fig:tiltDiffImages}.
The mirror changes can be evaluated with photometric apertures on key parts of mirror surfaces, such as in Figure \ref{fig:mirrorPhotTser} (right plot).
The time series of these apertures can cleanly sense the two tilt events like step functions as shown in Figure \ref{fig:mirrorPhotTser} (left second from the top).
We also calculate this photometry on the difference between pairs of adjacent detector groups to assess the mirror changes at a higher time cadence (colored points in Figure \ref{fig:mirrorPhotTser}).

The NIRCam grism time series and time series imaging modes are the only modes currently supported that allows simultaneous high cadence wavefront sensing along with the time series.
In some time series modes such as the NIRSpec Bright Object Time Series \citep{birkmann2022nirspecExoplanets} or MIRI low resolution time series \citep{kendrew2015LRSMIRI}, defocused imaging is not currently available for real-time wavefront sensing of the primary mirrors, so other measures are desirable to sense tilt events.
NIRISS Single Object Slitless mode (SOSS) \citep{doyon2012NIRISSFGS} has a de-focusing lens so this mode has some analogous capabilities as the short wavelength weak lens.

We explored some ways to sense tilt events with the NIRCam grism time series spectroscopy and the fine guidance sensor (FGS) fine guide data streams \citep{doyon2012NIRISSFGS}.
We experimented with a few different statistics for the FGS images.
We found that the flux from the telemetry stream\footnote{See the \texttt{SA\_ZFGINSTCT} keyword \url{https://jwst-docs.stsci.edu/methods-and-roadmaps/jwst-time-series-observations/jwst-time-series-observations-noise-sources}, \citep{jdox2016}} can sense the first tilt event (C6) but could not easily reveal the second tilt event (C1).
In addition, the first tilt event (C6) could be detected with a dot product with the average PSF or the $3 \times 3$ pixel flux of the time series but the second tilt event from the C1 mirror was harder to pull out of the noise.
The FWHM of the FGS 8$\times$8 images also shows step-like changes with the two tilt events but at low signal to noise.
We found that principal component analysis (PCA) could reveal evidence of both tilt events but is also not easily distinguishable above the noise.

The noise-weighted ``differential dot product'' was most effective in detecting tilt events with FGS.
\begin{equation}\label{eq:diffDotProduct}
D_i \equiv \left((\vec{A_i} - \vec{R})/ \vec{V} \right)\cdot \left(\vec{M} - \vec{R} \right),
\end{equation}
where $D_i$ is the differential dot product of 8$\times$8 images reshaped to 64-pixel linear vectors. In the expression, $\vec{A_i}$ is a single FGS \texttt{cal} data image, $\vec{R}$ is a reference image after the tilt event, $\vec{V}$ is the variance image, and $\vec{M}$ is a reference image before the tilt.
This method did use prior information about the timing of tilt events from NIRCam data to find the average image from before the tilt $\vec{M}$ and the average image after the tilt $\vec{R}$, but it may be possible to iteravely scan the time series to determine $\vec{M}$ and $\vec{R}$ from FGS alone.
This statistic essentially magnifies the differences between the shape of the PSF before and after the tilt event and weights by the noise.
Figure \ref{fig:mirrorPhotTser} shows that both tilt events can be sensed with FGS using a dot product, where the $\vec{M}$ and $\vec{R}$ are chosen for the first tilt event (C6) and second tilt event (C1).

Additionally, the full width at half maximum (FWHM, bottom left plot in Figure \ref{fig:mirrorPhotTser}) of the long wavelength grism PSF is sensitive to tilt events because the PSF changes with the mirror tilt.
Thus, even when weak lens imaging is not available, the FGS images and science instrument FWHM can be used in all time series modes to give clues about tilt events.
However, Equation \ref{eq:diffDotProduct} requires some knowledge of the timing to find $\vec{M}$ and $\vec{R}$ unless they can be found iteratively.
Therefore, a more general-purpose indicator of tilt events may be the FWHM of science data, which can be tracked in all time series observations.

Tilt events are hypothesized to be structural microdynamics in the telescope that may occur when stresses in the backplane structure behind the mirrors are suddenly relaxed.
As these stresses are released, the frequency of tilt events is expected to decrease in time.
Regular wavefront measurements indicate that they are less frequent in the transition from commissioning to science observations in July 2022 than earlier in commissioning.
However, tilt events were common even 4.5 months between launch and these observations, and similar tilt events may occur during Cycle 1 science observations.

\subsection{Tilt Timescale}\label{sec:tiltTimescale}

\begin{figure*}
\gridline{\fig{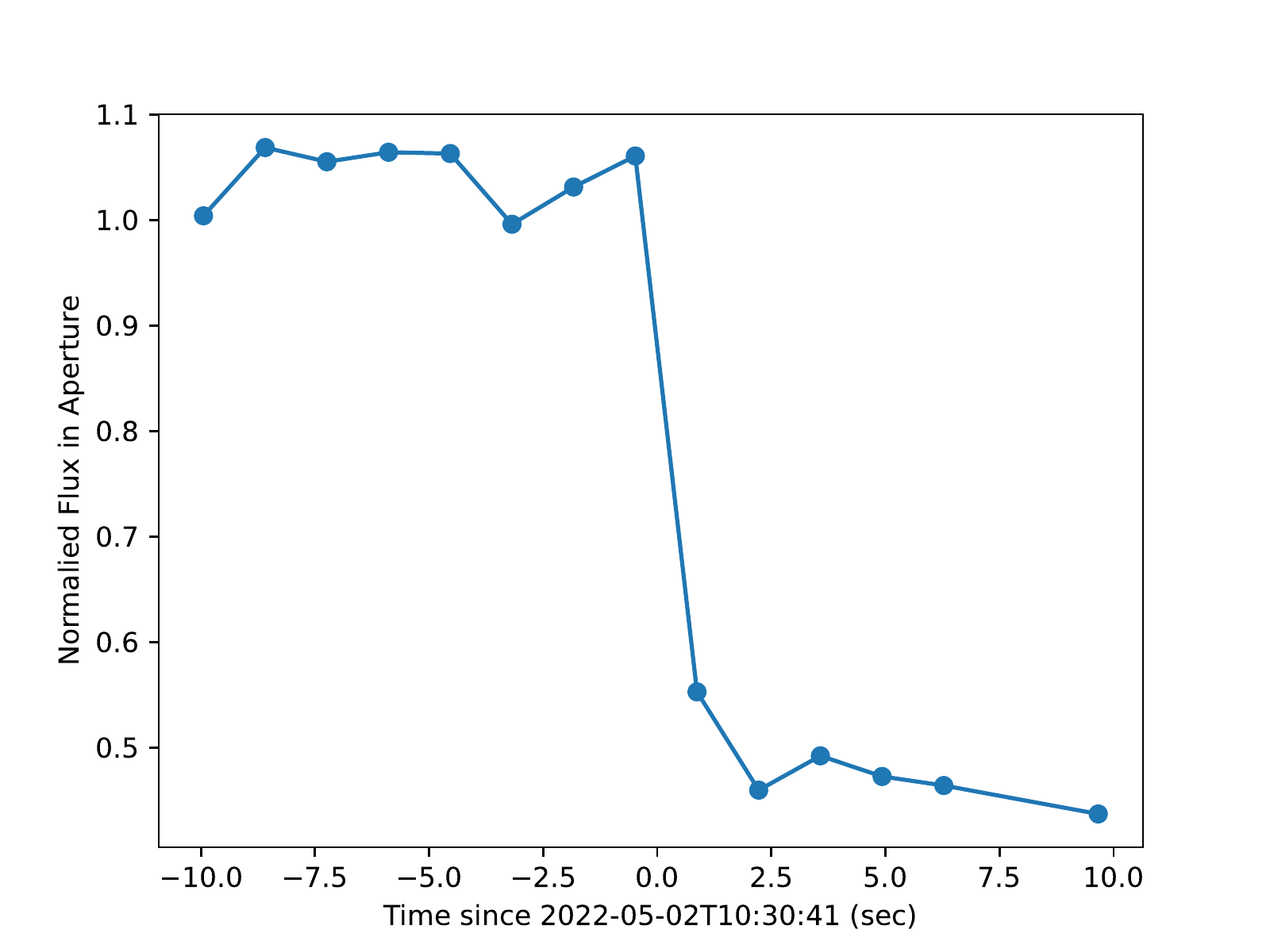}{0.48\textwidth}{Zoom-in on the C6 Tilt Event}
\fig{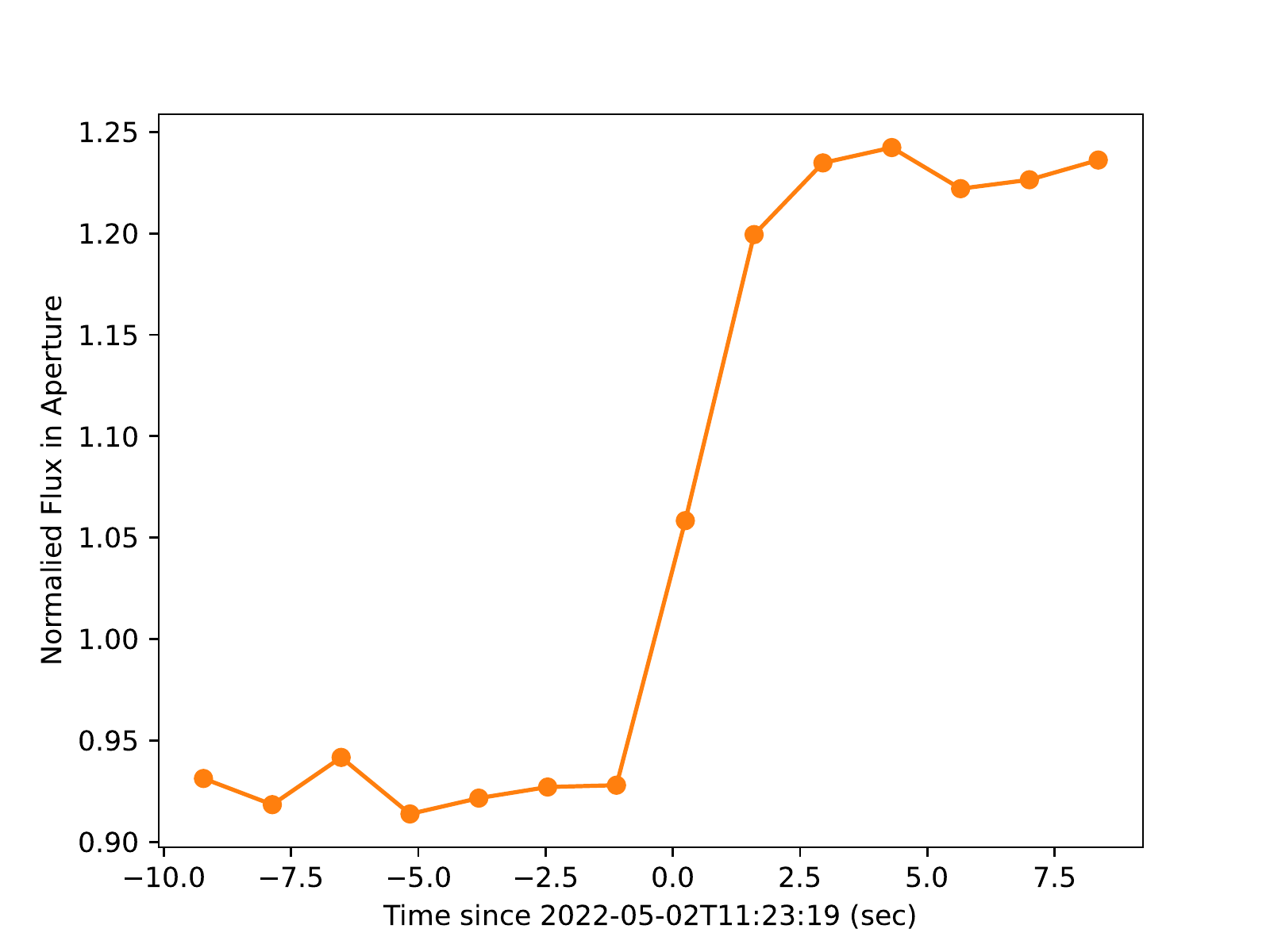}{0.48\textwidth}{Zoom-in on the C1 tilt event}}
\caption{Mirror photometry on the pairwise-subtracted images, zoomed in on events.
We use the same apertures as shown in Figure \ref{fig:mirrorPhotTser} to sense the tilts on the C1 and C6 primary mirror segments.
These two happen on a timescale as fast or faster than the 1.351 second time between BRIGHT2 detector groups, which contain 2 frames each.
For the C1 tilt event, the jump likely happened in the middle of a detector group so it has a sample between the low and high flux values.\label{fig:mirrorPhotTserZoom}}
\end{figure*}

The mirror changes appear to be step functions in Figure \ref{fig:mirrorPhotTser}, so the change happens on a timescale faster than the 27 second cadence of integrations.
We inspect the mirror changes at a higher cadence (but noisier) time series by calculating the mirror tilt specific photometry on the group-by-group difference images, which have a cadence of 1.351 seconds.
In this analysis, adjacent points are thus anti-correlated because they share a detector group so if one sample is anomalously high, the next one will be anomalously low.
As shown in Figure \ref{fig:mirrorPhotTserZoom}, there is a rapid change that takes 2 sampling durations of 1.351 seconds each to go from low to high or high to low, or in other words one sample between the low and high values.
There are two causes for the intermediate sample: 1) the tilt timescale is between 1.351 and 2.702 seconds and 2) the timescale is shorter than 1.351 seconds and occurs midway through the group.
We consider the first scenario less likely as it would require some fine-tuning to cause a tophat-like disturbance between these two time intervals that shows little sign of change in the PSF beyond 2.702 seconds.
We therefore expect that the timescale is faster than the shortest measurable time with the NIRCam photometry which is the 1.351 second group time.
We also note that these observations used the BRIGHT2 mode, which includes two frames within a group so the tilt could happen between these two frames within a group or within one of the frames.
We do not see any discontinuities within the image that would indicate a change within the frame, but the PSF only covers a fraction of the field of view so it is not a strong constraint.

\subsection{Charge Trapping Ramp}\label{sec:rampTimescale}
The HST WFC3 IR detector, which is an earlier generation HgCdTe detector, exhibits ramps with an exponential settling behavior \citep[e.g.][]{berta2012flat_gj1214,zhou2017chargeTrap}.
Ground-based laboratory tests showed likely negligible effects for typical JWST detectors \citep[e.g.][]{schlawin2020jwstNoiseFloor2}.
However, the NRCA3 short wavelength detector exhibits the most persistent charge and likely charge trap density of the 10 NIRCam detectors that are used in flight \citep{leisenring2016persistence}.
NRCA3 is the same detector used for the short wavelength component of the grism time-series observations when the long wavelength filter is selected to be F277W, F322W2, F356W.\footnote{\url{https://jwst-docs.stsci.edu/jwst-near-infrared-camera/nircam-operations/nircam-target-acquisition/nircam-grism-time-series-target-acquisition}}
The NRCA1 detector that is paired with the F444W filter, had about half the accumulated counts from charge trap release as the NRCA3 detector in ground based tests \citep{leisenring2016persistence}, so it is expected to have a smaller amplitude exponential ramp.

\begin{figure*}
\gridline{\fig{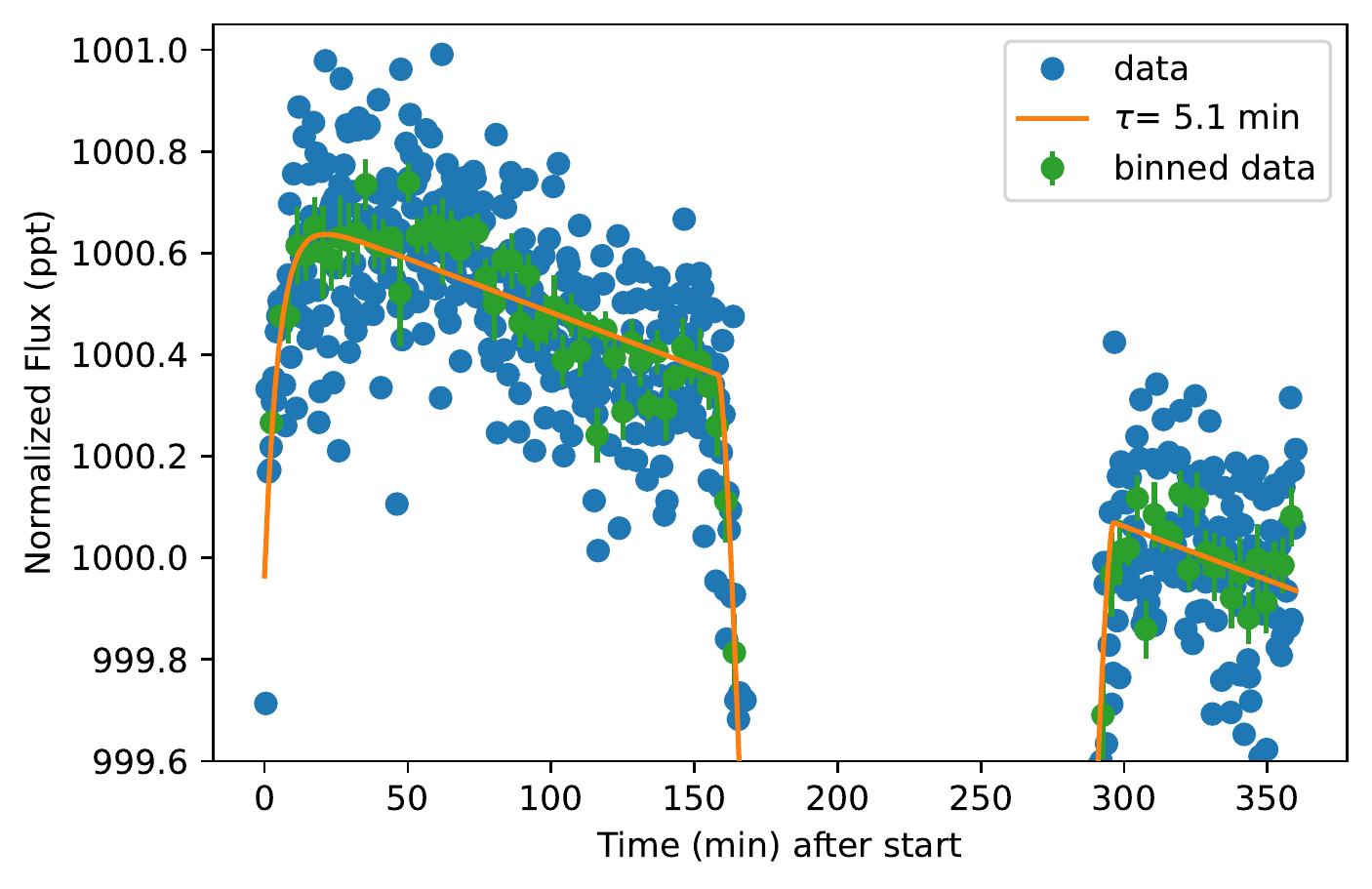}{0.48\textwidth}{Lightcurve Fit with a Linear Baseline}
\fig{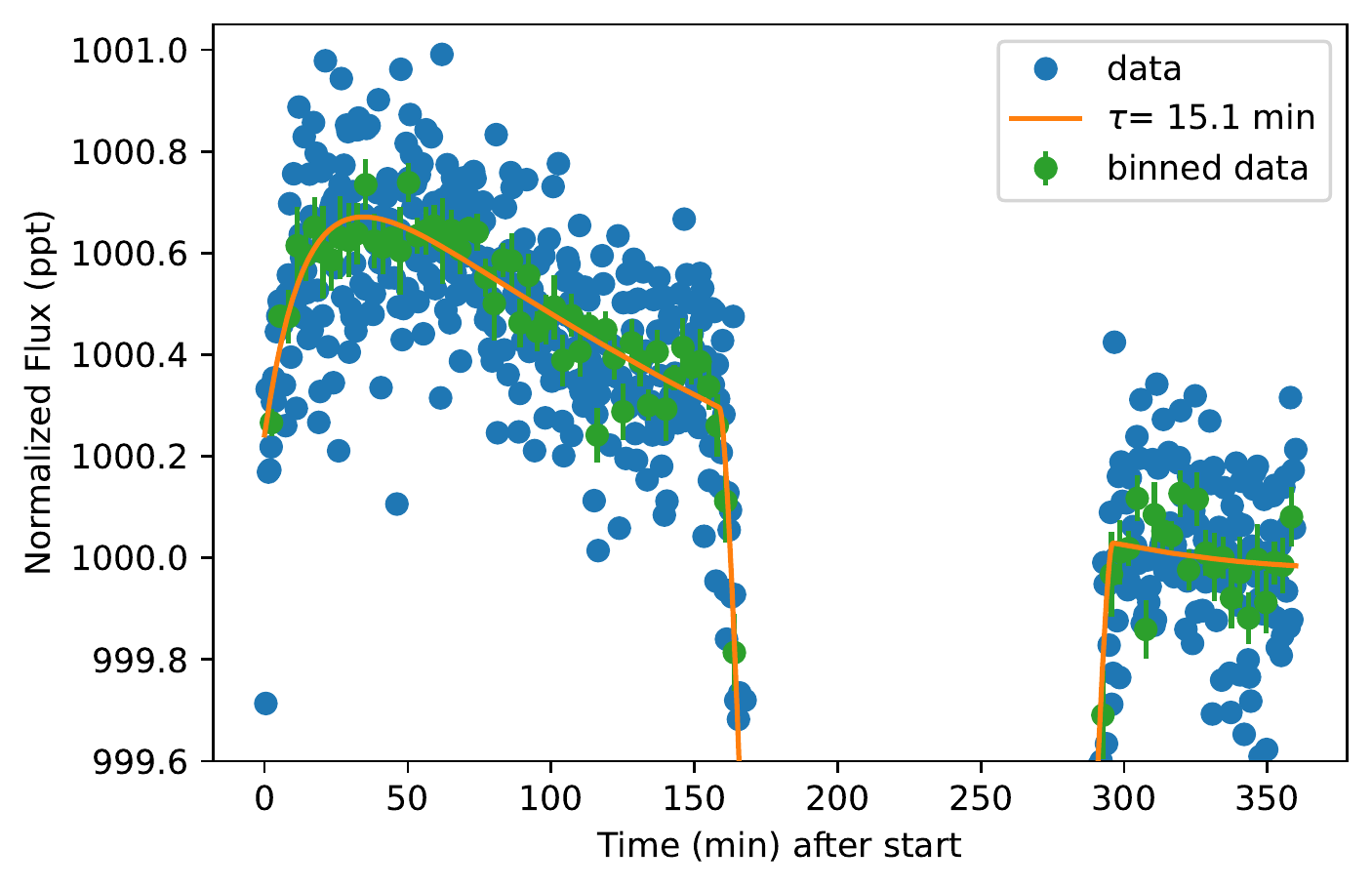}{0.48\textwidth}{Lightcurve Fit with a Quadratic Baseline}}
\caption{The lightcurve settling behavior is very fast with a fitted settling timescale of \settlingTime\ but is somewhat degenerate with the time series baseline.
}\label{fig:rampTimescale}
\end{figure*}

We fit an exponential model to the charge trapping behavior, as is common with HST charge trapping ramps \citep[e.g.][]{berta2012flat_gj1214} and show the results in Figure \ref{fig:rampTimescale}.
We first correct the small 260 ppm jump in the lightcurve using a wavefront model described in Section \ref{sec:tiltEvents}.
We fit the lightcurve with the following planet and systematics model: 
\begin{equation}
f(t) = \left(A + B x' + C x'^2\right) \left( 1- R \exp{(x - x_0)/\tau} \right) f_a(t),\label{eq:lightcurveModel}
\end{equation}
where A, B and C are coefficients in the quadratic baseline trend and normalization, $x$ is barycentric time, $R$ is the exponential amplitude, $x_0$ is the exposure start time, $f_a(x)$  is the astrophysical limb darkened lightcurve \citep[\texttt{starry},][]{luger2019starry}, with a 4 parameter polynomial limb darkening law from ExoCTK and $x'$ is the normalized time,
\begin{equation}\label{eq:expSettling}
x' = \frac{x - x_{med}}{x_{max} - x_{min}},
\end{equation}
where $x_{med}$ is the median time, $x_{max}$ is the maximum time and $x_{min}$ is the minimum time.
We fit the lightcurves by fixing $C=0$ (ie. a linear fit) and also letting it be a free parameter (ie a quadratic fit).
As will be shown in Section \ref{sec:photPerformance}, significant correlated errors are seen in the residuals if a linear baseline trend is adopted instead of a quadratic trend.

We find an exponential amplitude $R$ of 731 ppm and an exponential time $\tau$ of 5.1 minutes for a linear trend and $R$ of 656 ppm and $\tau=$15 of minutes for a quadratic trend.
The normalization and polynomial trend constants for the quadratic trend are A=${1000.24 \pm 0.01}$ parts per thousand (ppt), B=${-0.911 \pm 0.03}$ ppt and C=${0.80 \pm 0.14}$ ppt.
In absolute units, this corresponds to a slope of -0.15~ppt/hr and derivate of the slope of 0.022~(ppt/hr$^2$).
We note that initially, we fit the baseline and exponential start to just the initial part of the lightcurve and found an order-of-magnitude exponential settling timescale of 11 minutes \citep{rigby2022jwstPerformance}.
The 5.1 and 15 minutes minutes came from a MCMC Bayesian fit to the full lightcurve with a linear and quadratic baseline respectively, whereas the 11 minutes comes from least squared minimization of the first 300 points (before planet ingress) and has a steeper linear slope.
We have not determined the cause of the long timescale trend, but note a that different linear trends were seen on the NRS1 and NRS2 detectors \citep{espinoza2022hatp14Spec} so it may be detector-related.
The quadratic slope and exponential ramp terms in Equation \ref{eq:expSettling} are correlated and thus the change in slope is fit with a different exponential settling time.
We also looked through the previous JWST activities and found that NIRCam's previous use was 37 hours before for wavefront sensing.
NIRCam detectors remained in idle reset mode for those 37 hours so it is unlikely any long timescale traps were filled and are being released to create the downward slope seen in the data.

\subsection{Non-Linearity Effects}\label{sec:nonLinearEffects}

The H2RG detectors used in the near infrared instruments on JWST are never strictly linear at any well filling fraction, ranging from sub-percent non-linearity to tens of percent at 98\% the hard saturation value \citep[e.g.][]{canipe2014nonlinCorrection}.
However, they become increasingly non-linear as the detector approaches full well capacity and saturation \citep[e.g.][]{plazas2017nonlinearityAndPixelShifting}.
Correction polynomials are applied by the CalWebb JWST pipeline \citep{bushouse2022jwstPipeline} to linearize the counts as function of counts (DN).
A different but analogous method has been shown that linearity corrections are possible even up to high well filling fractions \citep[$\sim$97\%][]{canipe2014nonlinCorrection}.

\begin{figure*}
\gridline{\fig{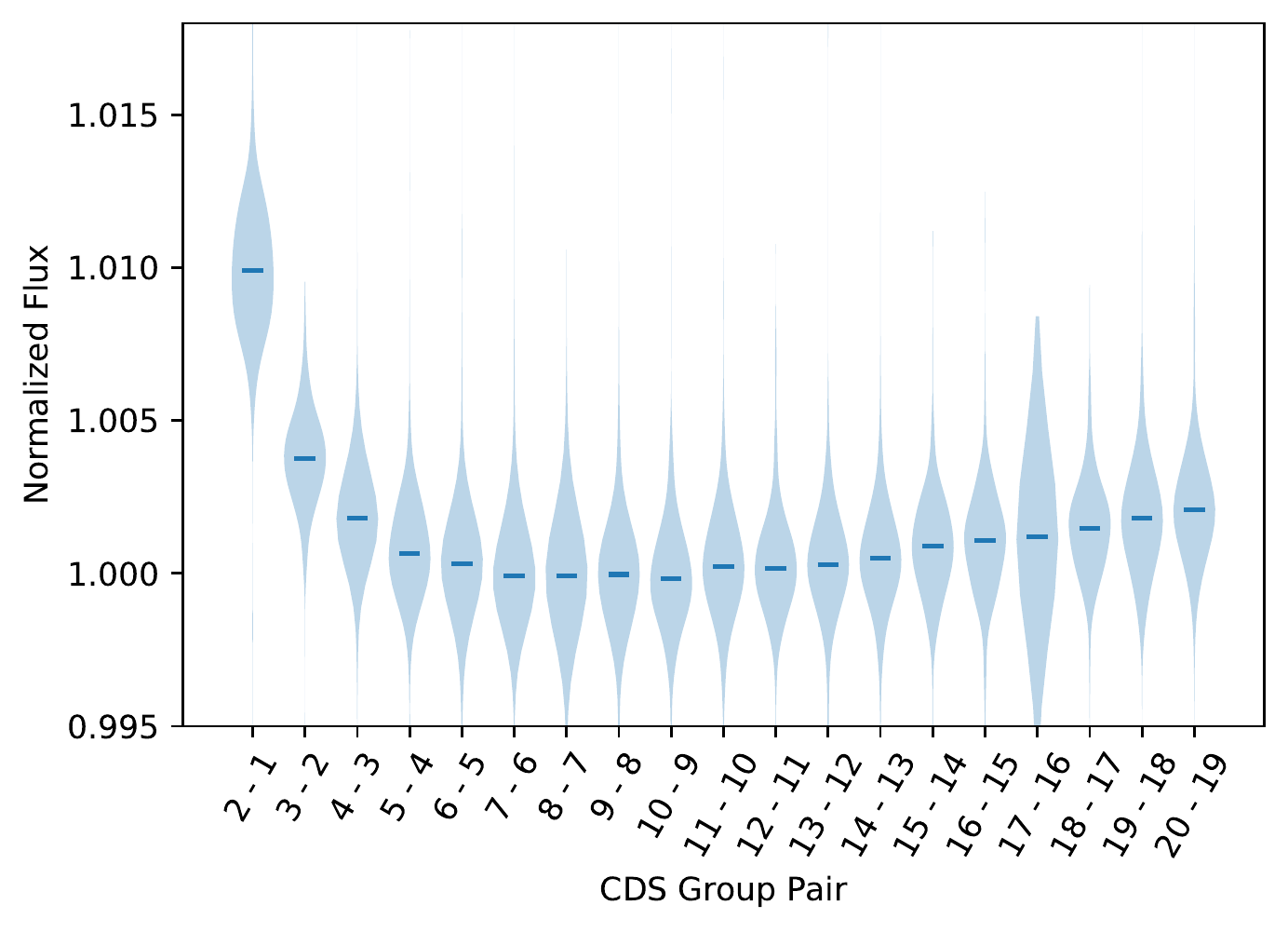}{0.7\textwidth}{}}
\caption{The difference in counts between groups up the ramp is not constant across an integration, as would be expected after linearity correction curves are applied.
The violin-shaped points are the distributions of fluxes as calculated by the pairwise difference image of two detector groups, using all of the out-of-transit integrations .
The median flux from all out-of-transit integrations is shown as a short horizontal line.
Future observations with just 2 or 3 total groups may exhibit absolute flux differences and transit depth differences due to this non-linearity soon after detector reset.
}\label{fig:nonLinEffects}
\end{figure*}

We assessed the difference between pairs of groups up the ramp to look for evidence of non-linearity.
The stellar flux is highly stable within a 27 second long integration so it is expected that a linearity-corrected ramp should have a constant difference between each successive group.
We find however that the first 4 group differences and the last 9 have a higher rate of flux than the middle 6 group differences, as shown in Figure \ref{fig:nonLinEffects}.
The early higher fluxes may be related to the release of trapped charge \citep[e.g.][]{smith2008imgPersistence,leisenring2016persistence} or a type of reset anomaly \citep{rauscher2007detectors}.
We do not perform a dark current subtraction, as discussed in Section \ref{sec:pipeAdjustments}.
However, the dark current is less than 0.05 e$^-$/sec on the short wavelength NIRCam detectors, and the 1\% change in the differential samplings (shown in Figure \ref{fig:nonLinEffects}) on top of a representative rate of 400 e$^-$/s would require 4 $e^-$/s of dark current.
So the reset anomaly seen here may be different from the anomaly seen on other H2RG devices \citep{rauscher2007detectors}.

Exoplanet time series observations with a small number of groups ($\lesssim 5$) could have an absolute photometric flux offsets but also systematic differences in the transit depth as compared to observations with many groups ($\gtrsim 5$).
This is because the non-linearity effects can change the response of the detector to a differential signal (ie. a non-constant derivative of count rate as a function of signal in e-/s/Mjy).
With the observed 1\% change in signal from group 2-1 to group 4-3 in Figure \ref{fig:nonLinEffects}, a 2\% deep transit of a hot Jupiter could have a measured transit depth of up to $\sim$ 1\% $\times$ 2\% $\times$ 1/4 = 50 ppm deeper if just using group 2 and 1.
This difference is below the measurement noise for a single transit of HAT-P-14 b using only group 2 - 1 but could be assessed on a brighter target.
There is also a possibility of reciprocity failure with HgCdTe detectors that can cause systematic changes in transit depth \citep{biesiadzinski2011reciprocityFailure,schlawin2020jwstNoiseFloor2}.

\section{Lightcurve Performance}\label{sec:NoiseAndStabilityPerformance}

\begin{figure*}
\gridline{\fig{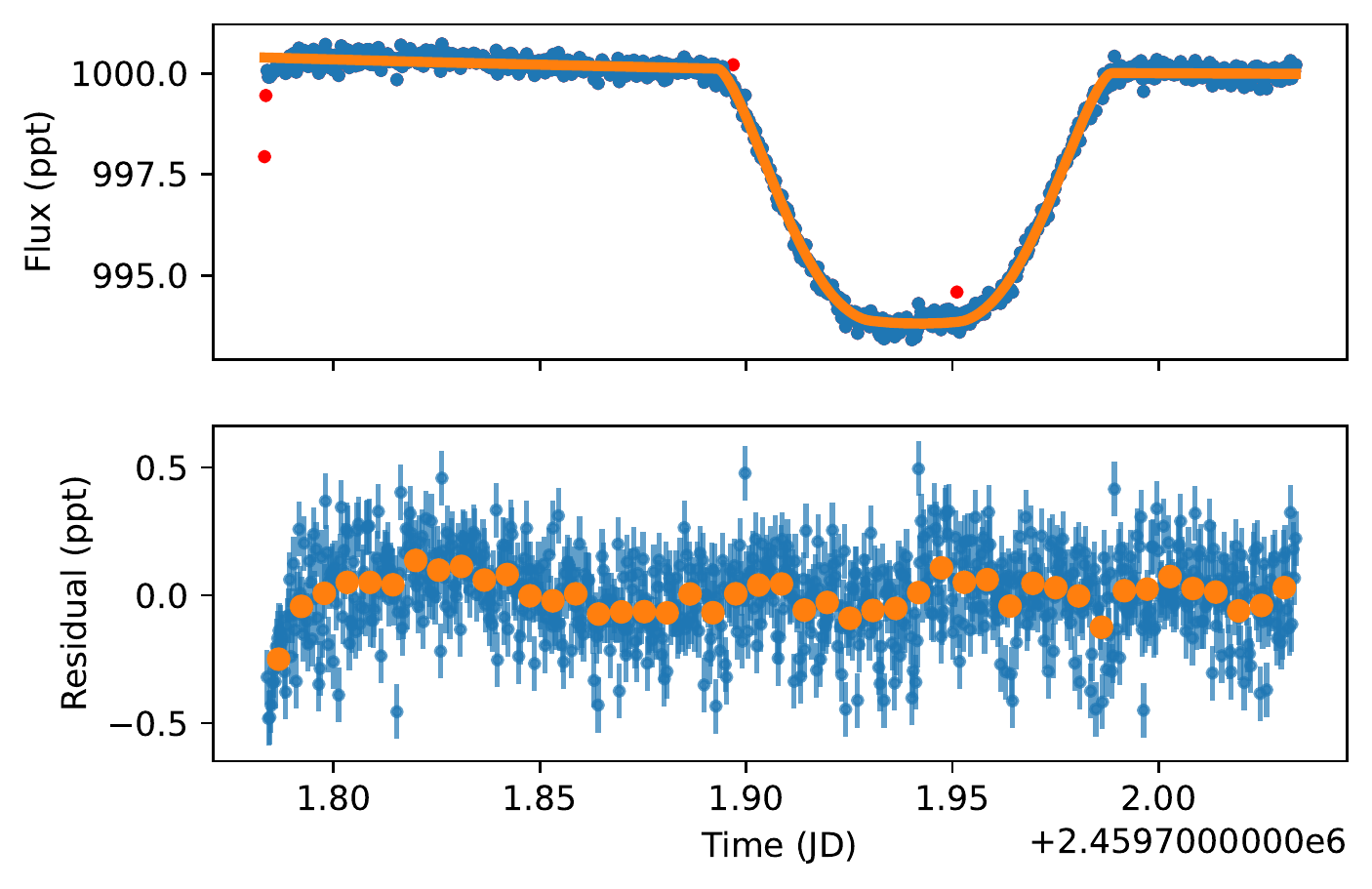}{0.48\textwidth}{Lightcurve fit with Quadratic Baseline Only}
\fig{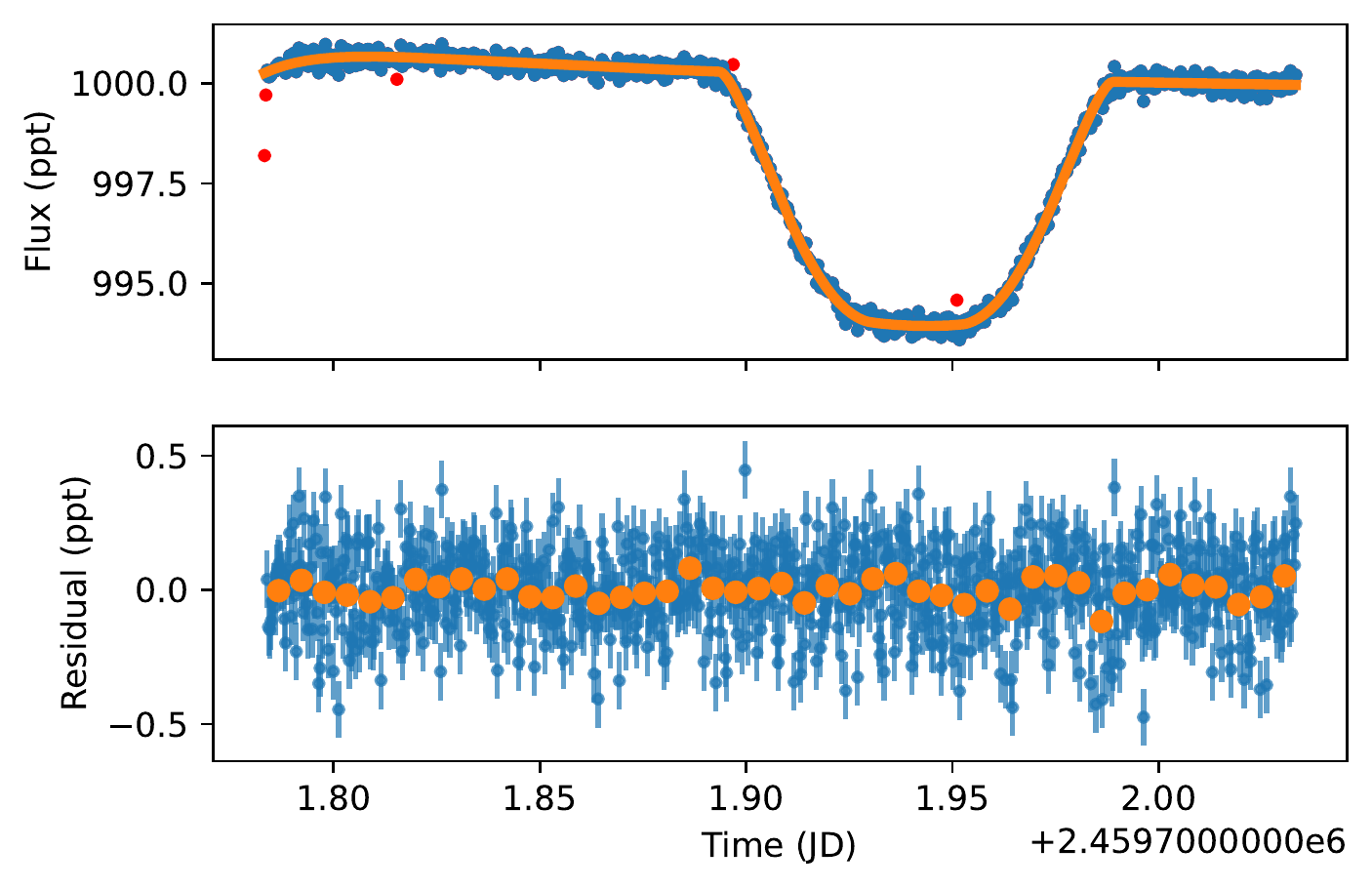}{0.48\textwidth}{Lightcurve with Jump and Exponential Ramp Model}}
\caption{Lightcurve and residuals before and after corrections for an exponential ramp and jump from the first tilt event.
The blue symbols with error bars are the fluxes for each integration at a cadence of 27.7 seconds, while the orange symbols are 8 minute long time-integrated bins that aid in viewing smaller-level changes.
Red symbols are outlier points not used in the fitting.\label{fig:lcAndResiduals}}
\end{figure*}

\subsection{Lightcurve Fitting}\label{sec:lcFitting}

We fit the lightcurve with a transit model to determine the lightcurve performance, discussed in Section \ref{sec:photPerformance} and the timing accuracy, discussed in Section \ref{sec:timeAccuracy}.
We remove a 260 ppm jump in the F210M (2.1 \micron) time series due to a tilt event using a wavefront model of the optics discussed in Section \ref{sec:tiltEvents}, allow a quadratic trend with time and fit an exponential settling ramp discussed in Section \ref{sec:rampTimescale}.
We use the integration mid-times in the barycentric reference frame and barycentric dynamical time standard \texttt{int\_mid\_BJD\_TDB} included in the \texttt{INT\_TIMES} extension of the JWST data products.

We fit the corrected lightcurve with a \texttt{starry} transit lightcurve model \citep{luger2019starry}, an exponential ramp, a quadratic trend in time and the 260 ppm jump correction.
The planet's near-grazing impact parameter \citep[0.91][]{fukui2016hatp14}) means that the planet does not traverse near the stellar midpoint and fitting the limb darkening parameters as free parameters can lead to large uncertainties in the planet's radius.
We fix the limb darkening law to a 6600~K, [Fe/H]=0.11, $\log{g}$ = 4.25 \citep{stassun2017gaiaRadiiMasses} ATLAS9 model \citep{kurucz2017atlas9} calculated with ExoCTK for the F210M filter for $\mu > 0.05$ where $\mu$ is the cosine of the angle between the line of signt and the emergent intensity \citep[e.g.][]{kipping2013uninfomativePriorsQuad} for a 4 parameter non-linear law.
We use the \texttt{starry} lightcurve model, which uses a polynomial limb darkening law \citep{agol2020exoplanetAnalytic}, so we fit the intensity function from a 4 parameter nonlinear law with a 6 parameter polynomial limb darkening law, which has been shown to be accurate to $\lesssim 0.5$ ppm differences in flux \citep{agol2020exoplanetAnalytic}.
We start with priors centered on the values from \citet{bonomo2017harpsMasses} for the period, inclination, eccentricity and argument of pericenter, but widened these in case of systematic errors to 83.5 $\pm 0.3\degree$, $e$=0.1071 $\pm$ 0.01 and $\omega$=106.1 $\pm 5\degree$.
We use the ephemeris from publicly vailable TESS data and a wide a/R$_*$ of 8.9 $\pm$ 1.0 centered on the value from \citet{stassun2017gaiaRadiiMasses}.
The resulting lightcurve fits and resiuals are shown in Figure \ref{fig:lcAndResiduals}.
We show a model fit that does not account for two of the systematics to better illustrate them as well as another that accounts for them (the charge trap ramp and tilt event jump).

\subsection{Photometric Performance}\label{sec:photPerformance}

After fitting the lightcurve with a transit model that includes a jump correction from a phase retrieval, an exponential charge trapping ramp and quadratic baseline described in Section \ref{sec:lcFitting}, we analyze the statistical properties of the residuals.
We include all points that are not marked as outliers (all points within 5 $\sigma$ of the best fit model) which includes 775 out of 780 total integrations.
The standard deviation (scatter) in all of the non-outlier residuals is 152~ppm, compared to a theoretical limit of 107 ppm from photon and read noise.
Thus, the measured noise is 42\% larger than the theoretical limit, some of which could be due to 1/f noise \citep{schlawin2020jwstNoiseFloorI}.
Even after ROEBA subtraction, there remains higher frequency (shorter timescale than the 5.24 ms row read time) noise.
We note that the background annulus subtraction reduced the standard deviation of out-of-transit flux in the lightcurve by 37\% over skipping the background annulus subtraction so there is likely residual 1/f noise.
We next assess time-correlations in the data by binning the residuals.

A key metric in the performance of the NIRCam photometry is how well the noise scales as a function of bin size.
This commonly is presented in the form of an Allan Variance plot \citep[e.g.][]{pont2006redNoiseTransits,croll2011wasp12}.
In this plot, the photometric scatter is computed as a standard deviation as a function of bin size.
In the case where each time sample is independent, the noise drops as $1/\sqrt{N}$ and this is plotted as ``white noise scaling''.
When the measured photometric scatter begins to diverge from the $1/\sqrt{N}$ power law, this is a sign that the time samples are correlated and time-dependent systematics are present.

\begin{figure*}
\gridline{
\fig{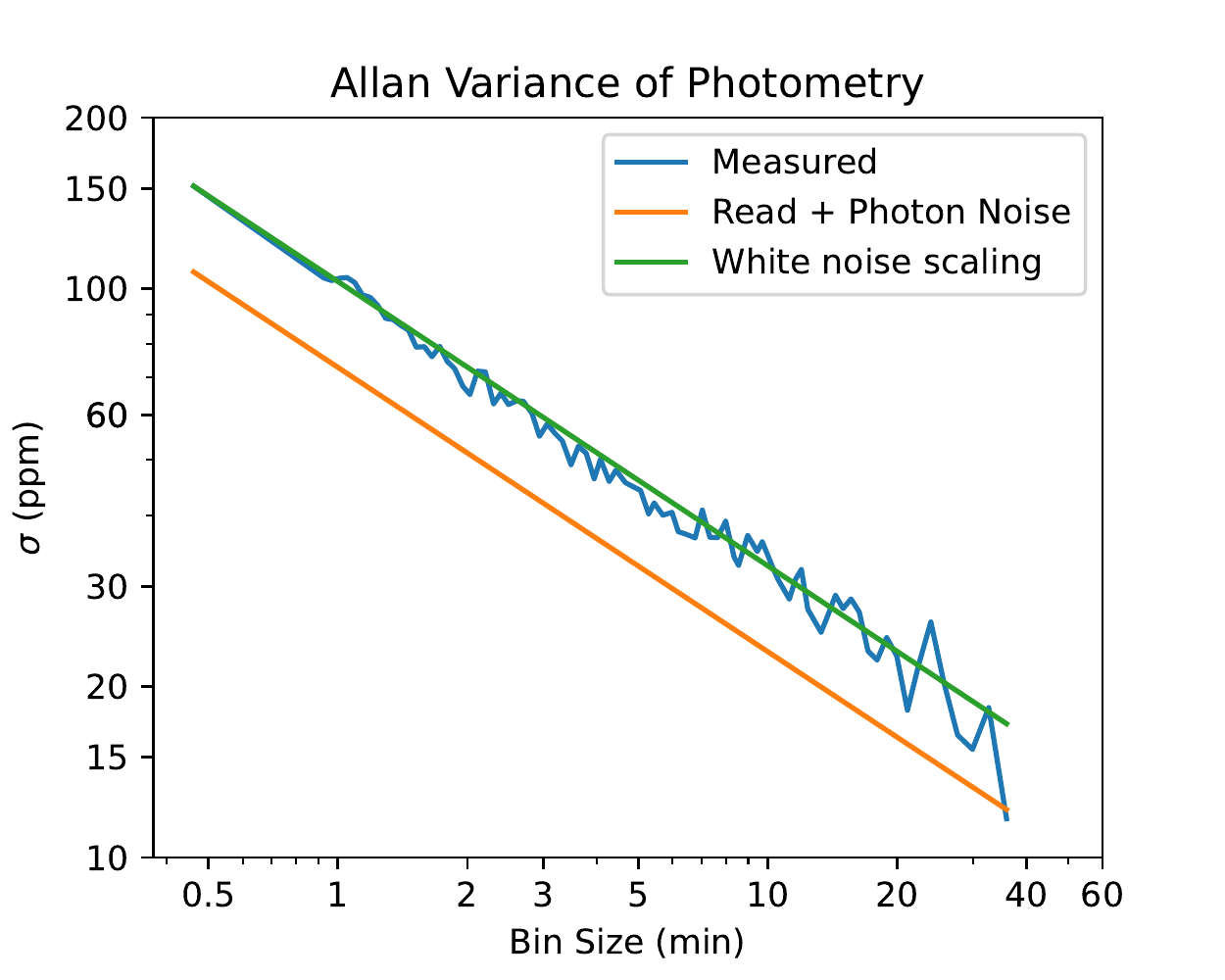}{0.48\textwidth}{Allan Variance of Quadratically-Corrected Lightcurve}
\fig{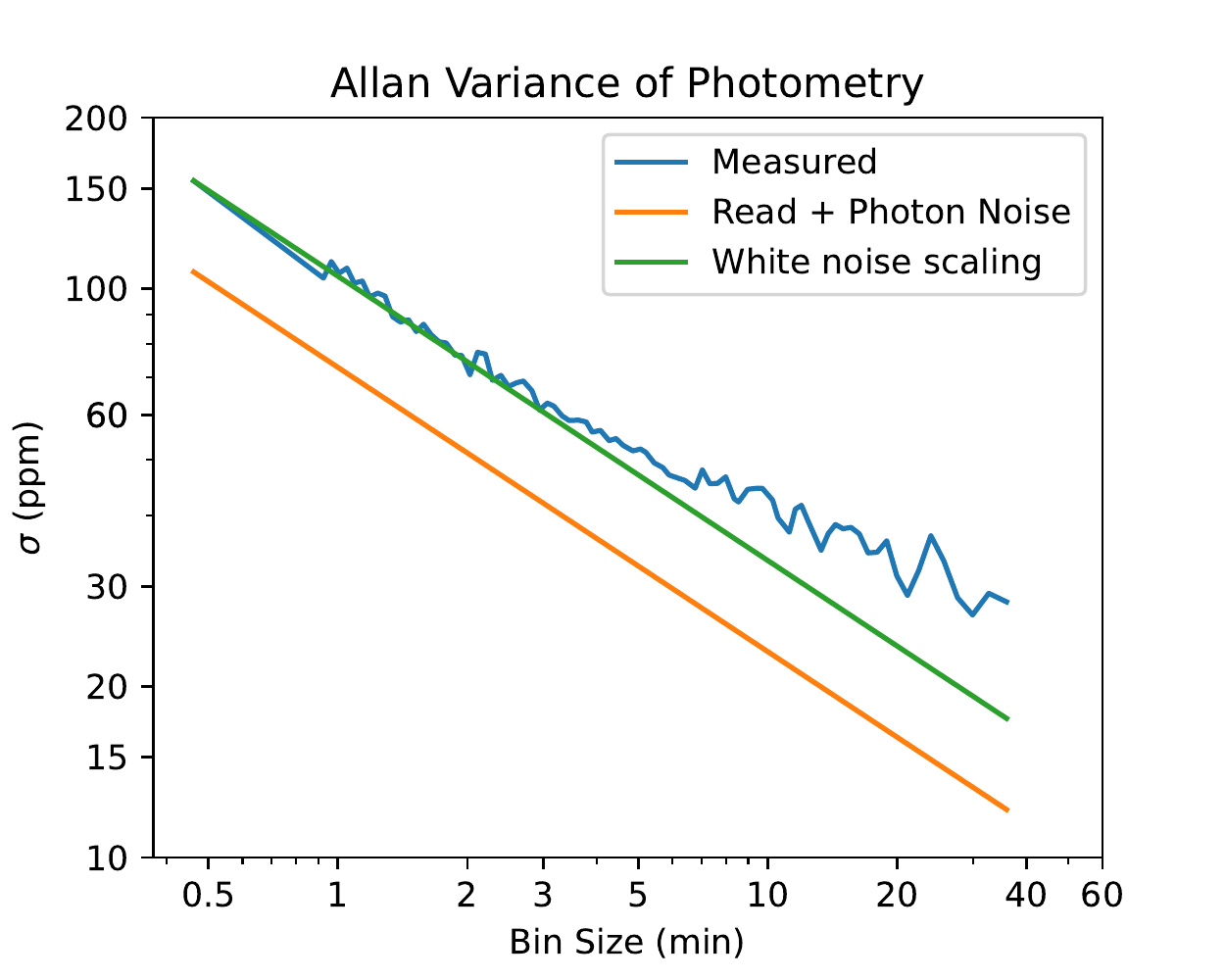}{0.48\textwidth}{Allan Variance of Linearly-Corrected Lightcurve}}

\caption{The lightcurve errors per bin drop nearly as $1/\sqrt{N}$ for $N$ time points, after correcting for the exponential startup, quadratic trend and transit (left plot curve).
Thus, the noise is largely independent for each integration.
If we only include a linear baseline trend to fit the lightcurves (right), there is excess noise beginning for bin sizes around 4 minutes.
The precision before time binning is about 42\% worse than the ideal limit of photon and read noise, likely because of 1/f noise correlations between pixels within each frame.}\label{fig:allanVar}
\end{figure*}

Figure \ref{fig:allanVar} (left) shows that the noise falls with $1/\sqrt{N}$.
Thus, no noise floor is measurable down to the 20 ppm level, but 1/f noise limits the precision in our current analysis to 42\% above the theoretical limit.
It may be possible to optimize the aperture to a hexagonal aperture, but previous tests with a circular annulus had little difference in noise performance on simulated data.
For this Allan variance analysis, we have removed the first tilt event from the lightcurve using a model discussed in Section \ref{sec:tiltEvents}, a quadratic trend with time and an exponential settling ramp discussed in Section \ref{sec:rampTimescale}.
For comparison, we also calculated an Allan variance curve when the baseline trend was fit with a linear polynomial instead of a quadratic polynomial in time.
In this curve, the noise begins to grow near 5 minutes where the noise is about 60 ppm.
Thus, there is some curvature to the lightcurve either related to instrument or astrophysical trends.

The measured 20 ppm lightcurve scatter at a time bin size of 30 minutes can be compared to the state-of-the-art best precision lightcurves from space-based photometers.
For the brightest targets observed by Kepler, the precision was measured to be $\sim$15 ppm \citep{jenkins2010initialCharKeplerLC}.
For the bright exoplanet system 55 Cnc, the precision with TESS was 10 ppm for 30 minutes \citep[][]{meierValdes2022TESS55Cnce} and for CHEOPS the precision achievable on a 1.6 hour eclipse observed 41 times using only out-of-eclipse data was 3~ppm \citep{demory2022CHEOPS55Cnce}, which would scale to 3 ppm * sqrt(1.6 hours/0.5 hours * 41 eclipses /2)=20~ppm for 30 minutes.
The scaling uses the relative time of the eclipse and the half hour benchmark, 41 separate eclipse events that were averaged and finally a factor of 2 accounting for the fact that the out-of-transit to in-transit ratio adds another factor of sqrt(2) in noise.
Elsewhere, the instrumental noise on hour-long timescales for CHEOPS is reported to be 15 to 80 ppm \citep{maxted2022pycheops}.
In the JWST photometry presented in this work, precision is limited by photon counting statistics, so brighter sources with more photons per minute will be needed to assess what is the highest possible precision as compared to the state-of-the-art performance.

\subsection{Pointing Performance}\label{sec:pointingPerformance}

Target acquisition was successful and placed the target at X=1060.7 px, Y=167.5 px on the NRCA3 detector in full frame 1-based Data Management System coordinates.
Pointing was stable to 0.01 pixels (0.3 mas) in the X direction and 0.009 px (0.3 mas) in the Y direction.
Some of the higher frequency jitter is thus averaged over the 27 second long integrations.
Thus, JWST attitude control provides incredible pointing stability at the 10$^{-2}$ pixel level and this produces no noticeable changes in flux with position and jitter, likely constituting a negligible part of the error budget.
The guide star has an FGS magnitude of 15.3 and we expect similarly high precision pointing performance for guide stars from magnitude 12.5 to 15.5.
Under the HgCdTe crosshatching structures on the NIRCam ALONG/A5 detector, 2 mas pointing was expected to produce 6 ppm changes of flux, so 0.3 mas jitter is likely to matter at the single digit ppm level and flux changes at this level cannot be measured for the HAT-P-14 target.

\subsection{High Gain Antenna Move}\label{sec:hgaMove}

\begin{figure*}
\gridline{\fig{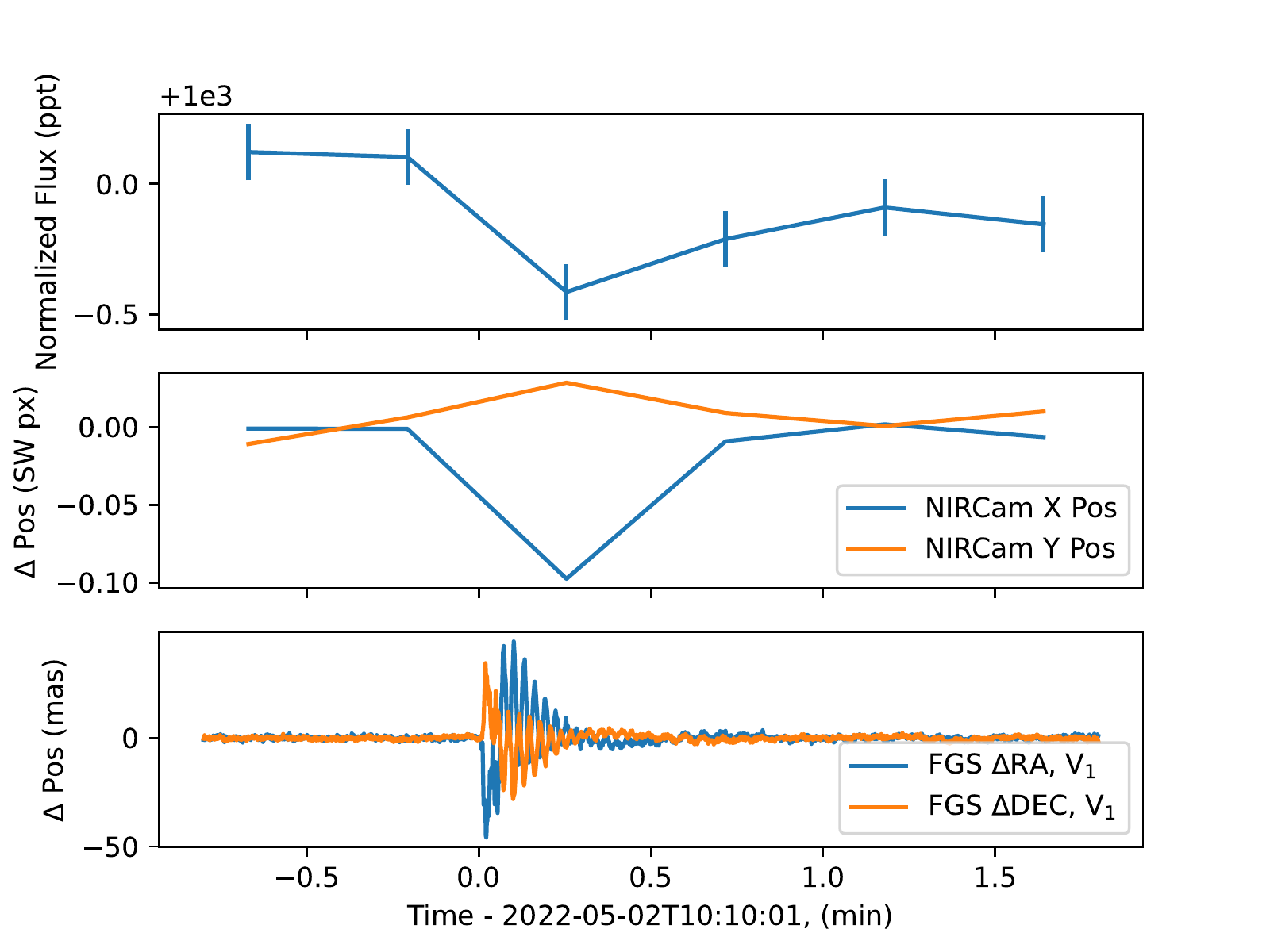}{0.49\textwidth}{Zoom-In on HGA move}
	    \fig{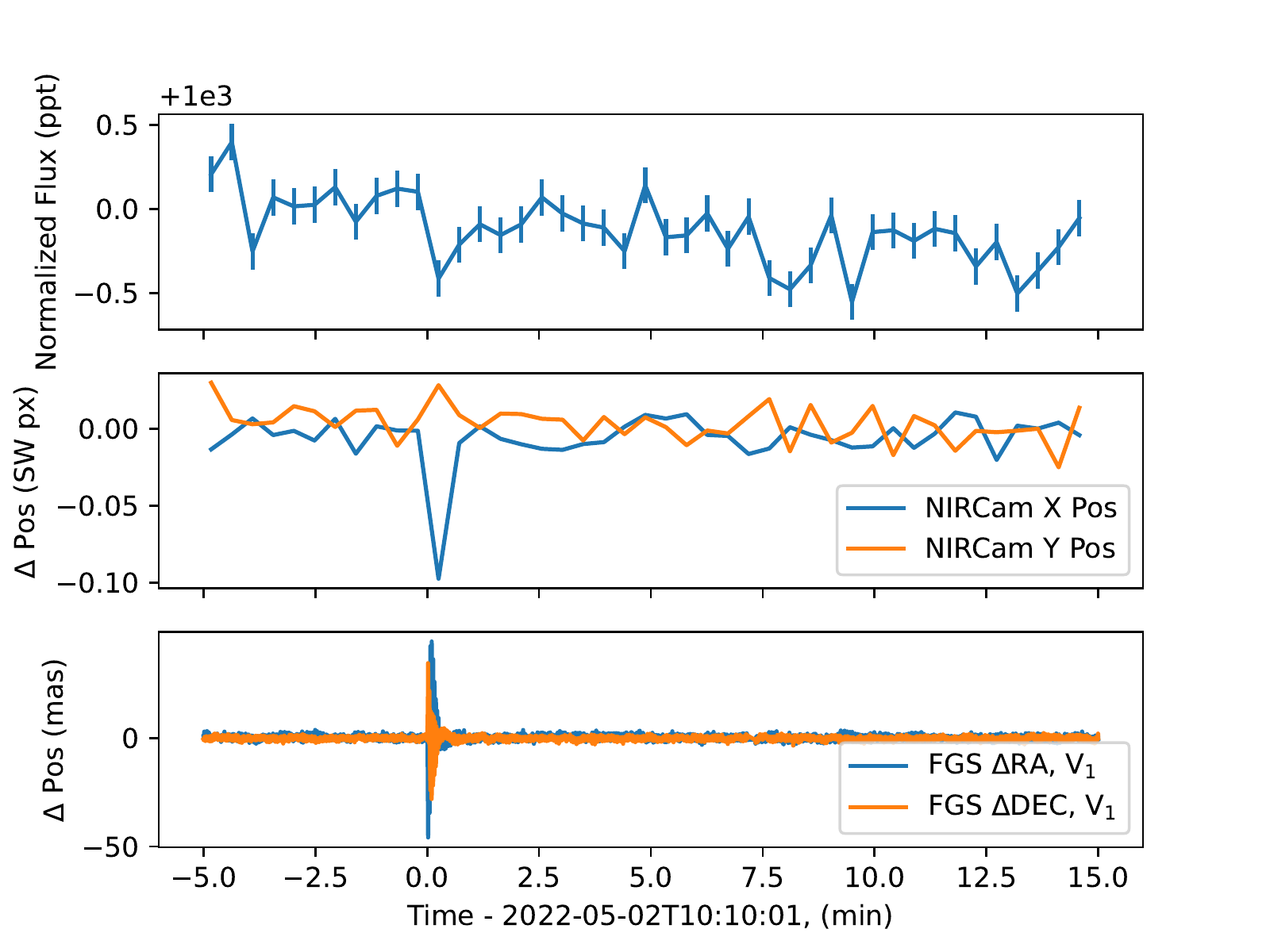}{0.49\textwidth}{Zoom-Out on HGA move}}

\caption{The High Gain Antenna moves are unlikely to have a big impact on time series observations.
The HGA move only affects about 0.5 minutes of data, which is mostly corrected by shifting the aperture but could be excluded from the lightcurve if it is an outlier.
The normalized flux changes by a maximum of about 0.5 parts per thousand (ppt).
The plate scale of NIRCam's Short Wavelength detector is 31 mas/px \label{fig:hgaMove}}.
\end{figure*}

JWST high gain antenna moves maintain pointing at the designated ground stations on Earth's Deep Space Network. While high gain antenna moves should happen during slews or between exposures, the moves are permitted to occur after 10,000 s during long time series observations with JWST. We had a high gain antenna adjustment at 2022-05-02T10:10:01 UTC, which was measurable with the Fine Guidance Sensor and NIRCam short wavelength time series centroids.
As shown in Figure \ref{fig:hgaMove}, the pointing change from the HGA move settles very quickly in less than 0.5 minutes.
Furthermore, the position was returned back to the original pointing to within 1 mas.
The data around an HGA move can be discarded and in the case of weak lens photometry that is spread over many pixels, only produced a transient 500 ppm change in flux.
Thus, HGA moves are not expected to cause significant issues to most time series observations as long as the Fine Guide mode does not lose a guide star from its subarray.

\subsection{Timing Accuracy}\label{sec:timeAccuracy}

We find a transit center time of 2022-05-02T10:34:55 +/- 7 seconds BJD$_{TDB}$, which is consistent (within 1.7 $\sigma$) with a prediction from the TESS ephemeris of 2022-05-02T10:36:35.1 +/- 60 seconds BJD$_{TDB}$.
Other planets with higher precision ephemerides will test JWST timing to higher accuracy.
There is also a dedicated GO program program identification number 1666 (PI Poshak Gandhi) that will use a double white dwarf binary and is expected to calibrate the absolute timing of an exposure to the 100 ms level.

\subsection{HAT-P-14 B Contamination}\label{sec:hatp14bContamination}
\begin{figure*}
\gridline{\fig{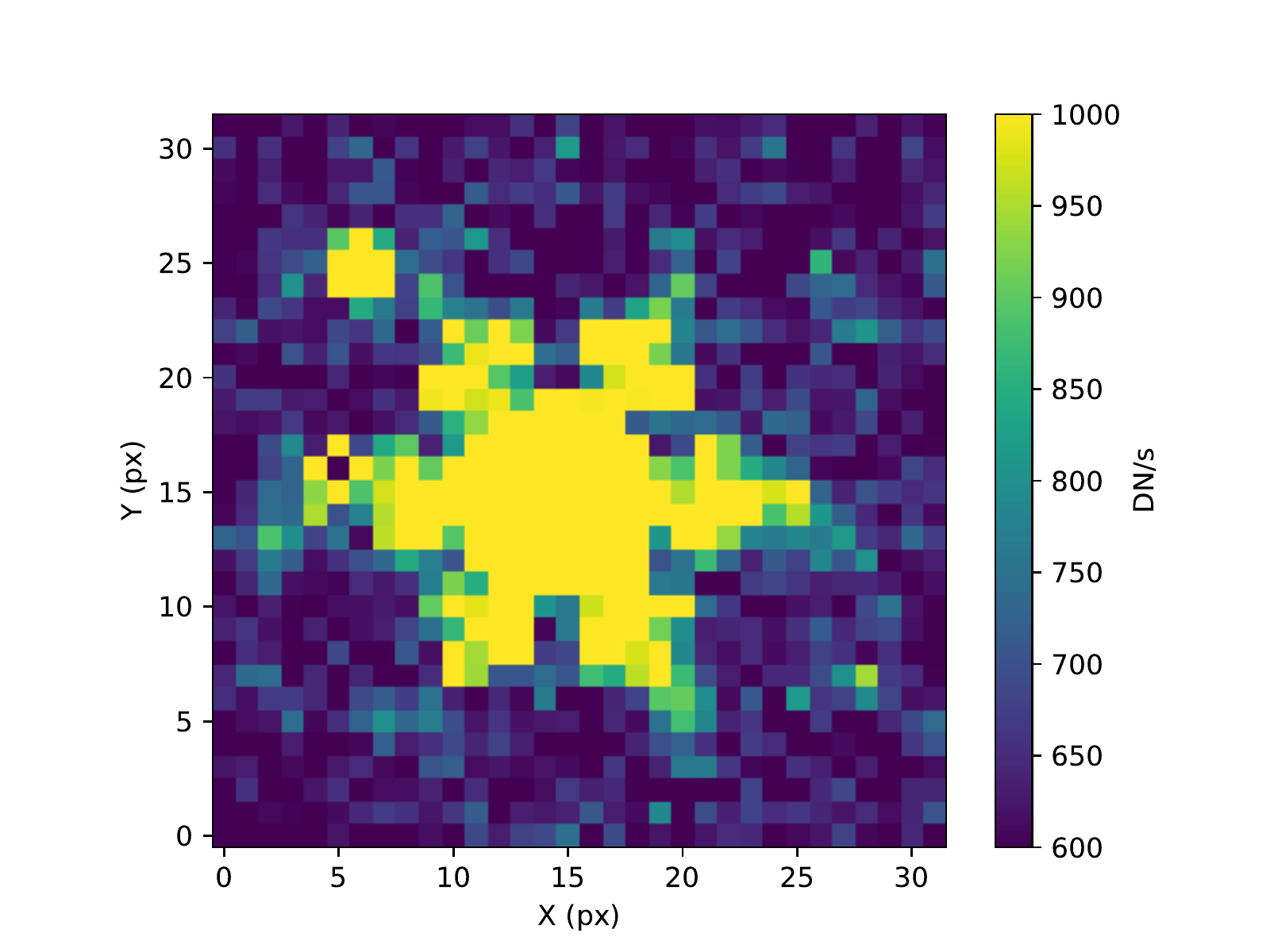}{0.4\textwidth}{}}
\caption{Long Wavelength Target Acquisition centered near HAT-P-14~A with a faint ($\Delta$mag=5) stellar companion HAT-P-14~B (upper left) at 860 mas separation that dilutes the transit depth.\label{fig:TAwithcompanion}}
\end{figure*}

The HAT-P-14 system was observed from the ground with Palomar adaptive optics (AO) on 2021-08-08 UTC to assess the contamination and dilution of the transit depth and find any unknown companions.
No companions were detected with a contrast less than $\Delta$mag = 7.0 within 0.5\arcsec.
The nearby stellar companion HAT-P-14 B \citep{ngo2015friendsHotJups} was also imaged with Palomar AO to better constrain the infrared colors and contamination on the transit depth.
The separation is 0.85\arcsec\ at a position angle of 264 degrees, which is well within the 2.5\arcsec\ source aperture used in the lightcurves of the central (hexagonal) portion of the PSF shown in Figure \ref{fig:wlp8PSF}.
With the Palomar AO imaging, we find a delta K magnitude between the HAT-P-14 A and B of $\Delta$K=4.99, which is a similar wavelength to this JWST F210M photometry.
The JWST target acquisition image shown in Figure \ref{fig:TAwithcompanion} shows both HAT-P-14 A and HAT-P-14 B.
Using the target acquisition image, we find that the contrast is $\Delta$ Mag$_{F335M}$=5.03 $\pm$ 0.05 for the F335M filter (3.35~\micron) and the separation is 0.83\arcsec.
This contrast is a similar to $K$ band, as expected for the Raleigh Jeans limit for long wavelengths.

Using \texttt{webbpsf}, we simulate the overlap of two Weak Lens +8 PSFs for a 81 pixel aperture and estimate the transit depth dilution.
We use the flight wavefront optical path difference evaluated at a time of 2022-05-02T10:30 UTC using \texttt{webbpsf} and perform aperture photometry on the simulated PSFs of HAT-P-14 A and HAT-P-14 B separately.
For the F210M filter at 2.1~\micron, we find that HAT-P-14 B should dilute the fitted 6540 $\pm$ 22 ppm transit depth by 59 ppm for an undiluted transit depth 6599 $\pm$ 22 ppm.
This is within 3.2~$\sigma$ of the long wavelength F322W2 broadband result of 6670 ppm, but with independent fitting methods and priors on orbital parameters and limb darkening.
This is also consistent with the NIRSpec time series broadband fit with 6627 $\pm$ 8 ppm with an independent analysis \citep{espinoza2022hatp14Spec}, with different priors on orbital parameters and limb darkening.
We do not expect the atmosphere to contribute much more than 15 ppm as described in Section \ref{sec:intro} for 0.9 atmospheric scale heights, but atmospheric gases could contribute at 1.5$\sigma$ level for 2 atmospheric scale heights of absorption.

\section{Conclusions}\label{sec:conclusions}

Transiting exoplanet science and other time-variability studies will benefit from JWST high precision time series photometry using the defocused photometry that is collected simultaneously with the grism time series mode.
Here, we present performance results from the hot Jupiter HAT-P-14 b that was observed to commission the instruments with a known flat transmission spectrum signal.
We detect the planet transit at high precision and find that the transit depth is consistent with the NIRSpec result to within 28 ppm $\pm$ 20 ppm as expected for this relatively high gravity planet.

The weak lens time series is particularly informative about small changes in the PSF as a result of wavefront variations, and we found two clear tilt events where two different hexagonal primary mirror segments quickly changed orientation with a timescale $\lesssim 1.4$ seconds.
Difference images clearly show the tilt events associated with specific mirror segments.
These tilt events can cause jumps in time series' signal levels but can be predicted effectively for short wavelengths using a wavefront model.
We also show how tilt events can be sensed with other instrument modes that do not use defocusing optics.
FGS differential images and the FWHM of the PSF can both reveal tilt events as step functions and will be good metrics to assess mirror stability.

The NRCA3 detector exhibited a charge trapping persistence ramp that is similar in shape to the ones seen on HST, but it settles out quickly on a \settlingTime\ decay timescale with an amplitude of 660 to 700 ppm and does not have a big impact on the time series.

The overall precision is very high and has a low scatter over the residuals of 152 ppm compared to the photon and read noise value of 107 ppm per 27 second integration, so it is only 41\% above the theoretical limit for this quiescent F-type $K$=8.85 star.
Some of this excess noise may be due to residual 1/f noise not corrected by the row-by-row subtraction.
The noise bins down approximately with the square root of the number of points in a time bin, indicating that systematic noise is minimal after the exponential and jump corrections and a quadratic polynomial as a function of time.
Overall, the prospects are good for high precision time series measurements of exoplanets and other astrophysical phenomena.
Defoucsed photometric monitoring will be valuable for measuring mirror tilt events and mitigating their impact on high precision measurements.

\acknowledgments

\section*{acknowledgements}
Funding for E Schlawin is provided by NASA's Goddard Space Flight Center.
T Greene acknowledges funding from the JWST project via NASA WBS 411672.05.05.02.02.
This research has made use of the SIMBAD database, operated at CDS, Strasbourg, France;
and NASA's Astrophysics Data System Bibliographic Services.
Part of the research was carried out at the Jet Propulsion Laboratory, California Institute of Technology, under a contract with the National Aeronautics and Space Administration (80NM0018D0004).
We respectfully acknowledge the University of Arizona is on the land and territories of Indigenous peoples. Today, Arizona is home to 22 federally recognized tribes, with Tucson being home to the O'odham and the Yaqui. Committed to diversity and inclusion, the University strives to build sustainable relationships with sovereign Native Nations and Indigenous communities through education offerings, partnerships, and community service.  
Thank you to Eddie Bergeron for sharing analysis on detecting tilt events with FGS.

%

\vspace{5mm}
\facilities{JWST(NIRCam), JWST(FGS), Palomar}


\software{astropy \citep{astropy2013}, 
          \texttt{photutils v0.3} \citep{bradley2016photutilsv0p3},
          \texttt{matplotlib} \citep{Hunter2007matplotlib},
          \texttt{numpy} \citep{vanderWalt2011numpy},
          \texttt{scipy} \citep{virtanen2020scipy},
          \texttt{starry} \citep{luger2019starry},
          \texttt{pynrc} \citep{leisenring2020pynrc0p8dev},
          \texttt{webbpsf} \citep{perrin2014webbpsf},
          \texttt{pysiaf} \citep{sahlmann2019pysiaf}
          \texttt{jwst} \citep{bushouse2022jwstPipeline}
           }



\appendix

\section{Subarray Positions}\label{sec:subarrayPos}

\begin{figure*}
\gridline{\fig{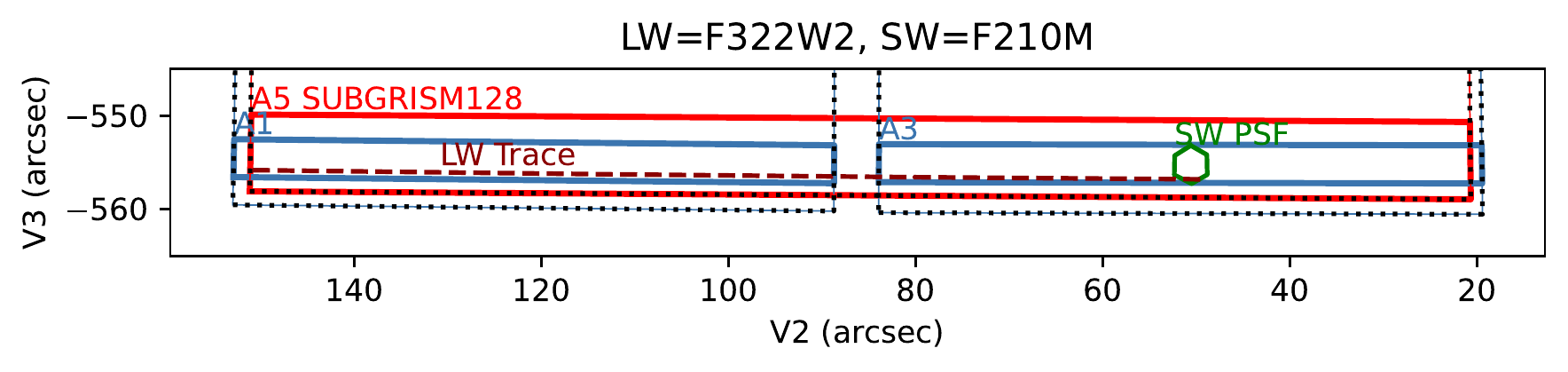}{0.99\textwidth}{}}
\gridline{\fig{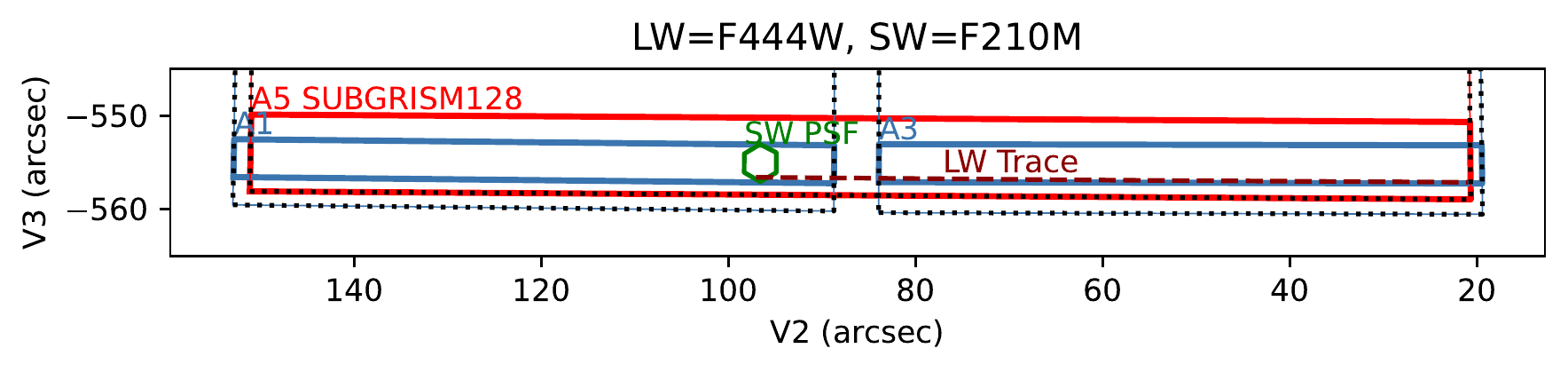}{0.99\textwidth}{}}
\caption{The NIRCam SW and LW channels observe approximately the same region of the sky simultaneously, however the overlap of detectors is not exact.
The relative position of the SW defocused weak lens image (green hexagon) is shown compared to the LW grism F322W2 (top figure) and F444W (bottom figure) central traces (brown dashed lines).
The SW SUBGRISM128 subarrays on the A1 and A3 detectors (blue solid lines) are aligned to capture light from the hexagonal +8 wave PSF while the the LW SUBGRISM128 subarray (solid red line) is aligned to capture light from the LW grism trace and surrounding background area.
The full frame detector boundaries (dotted lines) are where the 4 reference pixels are located, so the bottom references are included with the LW SUBGRISM128 subarray (solid red line) but no bottom reference pixels are available in the A1 and A3 SUBGRISM128 subarrays (solid blue lines).
All positions are shown in V2 and V3 Telescope coordinate system using \texttt{pysiaf} \citep{sahlmann2019pysiaf} after 8-10 SW pixel adjustments were made to the A1 and A3 subarrays following HAT-P-14 commissioning observations to better center the A1 and A3 subarrays on the hexagonal PSF.
 \label{fig:SWLWorientation}}
\end{figure*}

To help orient JWST users on the subarrays and relative positions of the short wavelength and long wavelength detector images, we provide diagrams of the locations as measured by in-flight data.
Figure \ref{fig:SWLWorientation} displays the relative locations, in telescope coordinates, of the NIRCam LW grism trace (dashed lines) and the NIRCam SW hexagonal PSF (green hexagon) from the +8 wave defocused pupil element (Weak Lens +8), which are observed simultaneously.
There is a small amount of refraction with the grism so there is a vertical offset between the two positions as placed in the same coordinate system.
The edges of the full frame images for the A1 (short wavelength), A3 (short wavelength) and A5 (long wavelength) detectors are all shown as dotted lines while the edges of the SUBGRISM128 subarrays for the same 3 detectors are shown as solid lines.
The positions here are shown for Cycle 1 after commissioning where small (8-10 SW px) tweaks were made to better-center the A1 and A3 subarrays on the hexagonal PSF.
The reference pixels on JWST NIRCam detectors are at the full frame boundaries, so the relative positions of the both SUBGRISM64 and SUBGRISM128 subarrays necessitate excluding the bottom 4 reference pixel rows on the short wavelength NRCA1 and NRCA3 detectors.
Significant improvements in performance can be made if background pixels are used in the same manner as reference pixels to correct for amplifier offsets and 1/f noise, as discussed in Section \ref{sec:pipeAdjustments}.

\bibliographystyle{apj}
\bibliography{this_biblio}

\begin{thebibliography}{}
\expandafter\ifx\csname adsurllinklabel\endcsname\relax
  \def\adsurllinklabel{[LINK]}\fi
\expandafter\ifx\csname adsadsurllinklabel\endcsname\relax
  \def\adsadsurllinklabel{[ADS]}\fi
\expandafter\ifx\csname natexlab\endcsname\relax\def\natexlab#1{#1}\fi

\bibitem[{{Agol} {et~al.}(2020){Agol}, {Luger}, \&
  {Foreman-Mackey}}]{agol2020exoplanetAnalytic}
{Agol}, E., {Luger}, R., \& {Foreman-Mackey}, D. 2020, \aj, 159, 123
 \href{https://ui.adsabs.harvard.edu/abs/2020AJ....159..123A}{\adsadsurllinklabel}

\bibitem[{{Ahrer} {et~al.}(2022){Ahrer}, {Stevenson}, {Mansfield}, {Moran},
  {Brande}, {Morello}, {Murray}, {Nikolov}, {Petit dit de la Roche},
  {Schlawin}, {Wheatley}, {Zieba}, {Batalha}, {Damiano}, {Goyal}, {Lendl},
  {Lothringer}, {Mukherjee}, {Ohno}, {Batalha}, {Battley}, {Bean}, {Beatty},
  {Benneke}, {Berta-Thompson}, {Carter}, {Cubillos}, {Daylan}, {Espinoza},
  {Gao}, {Gibson}, {Gill}, {Harrington}, {Hu}, {Kreidberg}, {Lewis}, {Line},
  {L{\'o}pez-Morales}, {Parmentier}, {Powell}, {Sing}, {Tsai}, {Wakeford},
  {Welbanks}, {Alam}, {Alderson}, {Allen}, {Anderson}, {Barstow}, {Bayliss},
  {Bell}, {Blecic}, {Bryant}, {Burleigh}, {Carone}, {Casewell}, {Changeat},
  {Chubb}, {Crossfield}, {Crouzet}, {Decin}, {D{\'e}sert}, {Feinstein},
  {Flagg}, {Fortney}, {Gizis}, {Heng}, {Iro}, {Kempton}, {Kendrew}, {Kirk},
  {Knutson}, {Komacek}, {Lagage}, {Leconte}, {Lustig-Yaeger}, {MacDonald},
  {Mancini}, {May}, {Mayne}, {Miguel}, {Mikal-Evans}, {Molaverdikhani},
  {Palle}, {Piaulet}, {Rackham}, {Redfield}, {Rogers}, {Roy}, {Rustamkulov},
  {Shkolnik}, {Sotzen}, {Taylor}, {Tremblin}, {Tucker}, {Turner}, {de
  Val-Borro}, {Venot}, \& {Zhang}}]{ahrer2022WASP39bERS}
{Ahrer}, E.-M., {Stevenson}, K.~B., {Mansfield}, M., {et~al.} 2022, arXiv
  e-prints, arXiv:2211.10489
 \href{https://ui.adsabs.harvard.edu/abs/2022arXiv221110489A}{\adsadsurllinklabel}

\bibitem[{{Astropy Collaboration} {et~al.}(2013){Astropy Collaboration},
  {Robitaille}, {Tollerud}, {Greenfield}, {Droettboom}, {Bray}, {Aldcroft},
  {Davis}, {Ginsburg}, {Price-Whelan}, {Kerzendorf}, {Conley}, {Crighton},
  {Barbary}, {Muna}, {Ferguson}, {Grollier}, {Parikh}, {Nair}, {Unther},
  {Deil}, {Woillez}, {Conseil}, {Kramer}, {Turner}, {Singer}, {Fox}, {Weaver},
  {Zabalza}, {Edwards}, {Azalee Bostroem}, {Burke}, {Casey}, {Crawford},
  {Dencheva}, {Ely}, {Jenness}, {Labrie}, {Lim}, {Pierfederici}, {Pontzen},
  {Ptak}, {Refsdal}, {Servillat}, \& {Streicher}}]{astropy2013}
{Astropy Collaboration}, {Robitaille}, T.~P., {Tollerud}, E.~J., {et~al.} 2013,
  \aap, 558, A33
 \href{http://adsabs.harvard.edu/abs/2013A%26A...558A..33A}{\adsadsurllinklabel}

\bibitem[{{Barstow} \& {Irwin}(2016)}]{barstow2016trappist1habitable}
{Barstow}, J.~K., \& {Irwin}, P.~G.~J. 2016, \mnras, 461, L92
 \href{http://adsabs.harvard.edu/abs/2016MNRAS.461L..92B}{\adsadsurllinklabel}

\bibitem[{{Bean} {et~al.}(2018){Bean}, {Stevenson}, {Batalha},
  {Berta-Thompson}, {Kreidberg}, {Crouzet}, {Benneke}, {Line}, {Sing},
  {Wakeford}, {Knutson}, {Kempton}, {D{\'e}sert}, {Crossfield}, {Batalha}, {de
  Wit}, {Parmentier}, {Harrington}, {Moses}, {Lopez-Morales}, {Alam}, {Blecic},
  {Bruno}, {Carter}, {Chapman}, {Decin}, {Dragomir}, {Evans}, {Fortney},
  {Fraine}, {Gao}, {Garc{\'\i}a Mu{\~n}oz}, {Gibson}, {Goyal}, {Heng}, {Hu},
  {Kendrew}, {Kilpatrick}, {Krick}, {Lagage}, {Lendl}, {Louden}, {Madhusudhan},
  {Mandell}, {Mansfield}, {May}, {Morello}, {Morley}, {Nikolov}, {Redfield},
  {Roberts}, {Schlawin}, {Spake}, {Todorov}, {Tsiaras}, {Venot}, {Waalkes},
  {Wheatley}, {Zellem}, {Angerhausen}, {Barrado}, {Carone}, {Casewell},
  {Cubillos}, {Damiano}, {de Val-Borro}, {Drummond}, {Edwards}, {Endl},
  {Espinoza}, {France}, {Gizis}, {Greene}, {Henning}, {Hong}, {Ingalls}, {Iro},
  {Irwin}, {Kataria}, {Lahuis}, {Leconte}, {Lillo-Box}, {Lines}, {Lothringer},
  {Mancini}, {Marchis}, {Mayne}, {Palle}, {Rauscher}, {Roudier}, {Shkolnik},
  {Southworth}, {Swain}, {Taylor}, {Teske}, {Tinetti}, {Tremblin}, {Tucker},
  {van Boekel}, {Waldmann}, {Weaver}, \& {Zingales}}]{bean2018ers}
{Bean}, J.~L., {Stevenson}, K.~B., {Batalha}, N.~M., {et~al.} 2018,
  Publications of the Astronomical Society of the Pacific, 130, 114402
 \href{https://ui.adsabs.harvard.edu/#abs/2018PASP..130k4402B}{\adsadsurllinklabel}

\bibitem[{{Beatty} {et~al.}(2022){Beatty}, {Schlawin}, {Greene}, {Brooks},
  {Espinoza}, N., J., M., M., B., D., A., S., A., M., M., D., \&
  M.}]{beatty2022hatp14Spec}
{Beatty}, T., {Schlawin}, E., {Greene}, T., {et~al.} 2022, in prep


\bibitem[{{Beichman} {et~al.}(2014){Beichman}, {Benneke}, {Knutson}, {Smith},
  {Lagage}, {Dressing}, {Latham}, {Lunine}, {Birkmann}, {Ferruit}, {Giardino},
  {Kempton}, {Carey}, {Krick}, {Deroo}, {Mandell}, {Ressler}, {Shporer},
  {Swain}, {Vasisht}, {Ricker}, {Bouwman}, {Crossfield}, {Greene}, {Howell},
  {Christiansen}, {Ciardi}, {Clampin}, {Greenhouse}, {Sozzetti}, {Goudfrooij},
  {Hines}, {Keyes}, {Lee}, {McCullough}, {Robberto}, {Stansberry}, {Valenti},
  {Rieke}, {Rieke}, {Fortney}, {Bean}, {Kreidberg}, {Ehrenreich}, {Deming},
  {Albert}, {Doyon}, \& {Sing}}]{beichman2014pasp}
{Beichman}, C., {Benneke}, B., {Knutson}, H., {et~al.} 2014, \pasp, 126, 1134
 \href{http://adsabs.harvard.edu/abs/2014PASP..126.1134B}{\adsadsurllinklabel}

\bibitem[{{Berta} {et~al.}(2012){Berta}, {Charbonneau}, {D{\'e}sert},
  {Miller-Ricci Kempton}, {McCullough}, {Burke}, {Fortney}, {Irwin}, {Nutzman},
  \& {Homeier}}]{berta2012flat_gj1214}
{Berta}, Z.~K., {Charbonneau}, D., {D{\'e}sert}, J.-M., {et~al.} 2012, \apj,
  747, 35
 \href{http://adsabs.harvard.edu/abs/2012ApJ...747...35B}{\adsadsurllinklabel}

\bibitem[{{Biesiadzinski} {et~al.}(2011){Biesiadzinski}, {Lorenzon}, {Newman},
  {Schubnell}, {Tarl{\'e}}, \&
  {Weaverdyck}}]{biesiadzinski2011reciprocityFailure}
{Biesiadzinski}, T., {Lorenzon}, W., {Newman}, R., {et~al.} 2011, \pasp, 123,
  958
 \href{https://ui.adsabs.harvard.edu/abs/2011PASP..123..958B}{\adsadsurllinklabel}

\bibitem[{{Birkmann} {et~al.}(2022){Birkmann}, {Ferruit}, {Giardino},
  {Nielsen}, {Garc{\'\i}a Mu{\~n}oz}, {Kendrew}, {Rauscher}, {Beck}, {Keyes},
  {Valenti}, {Jakobsen}, {Dorner}, {Alves de Oliveira}, {Arribas}, {B{\"o}ker},
  {Bunker}, {Charlot}, {de Marchi}, {Kumari}, {L{\'o}pez-Caniego},
  {L{\"u}tzgendorf}, {Maiolino}, {Manjavacas}, {Marston}, {Moseley}, {Prizkal},
  {Proffitt}, {Rawle}, {Rix}, {te Plate}, {Sabbi}, {Sirianni}, {Willott}, \&
  {Zeidler}}]{birkmann2022nirspecExoplanets}
{Birkmann}, S.~M., {Ferruit}, P., {Giardino}, G., {et~al.} 2022, \aap, 661, A83
 \href{https://ui.adsabs.harvard.edu/abs/2022A&A...661A..83B}{\adsadsurllinklabel}

\bibitem[{{Bitsch} {et~al.}(2021){Bitsch}, {Raymond}, {Buchhave},
  {Bello-Arufe}, {Rathcke}, \& {Schneider}}]{bitsch2021dryOrWaterSubNeptune}
{Bitsch}, B., {Raymond}, S.~N., {Buchhave}, L.~A., {et~al.} 2021, \aap, 649, L5
 \href{https://ui.adsabs.harvard.edu/abs/2021A&A...649L...5B}{\adsadsurllinklabel}

\bibitem[{{Bonomo} {et~al.}(2017){Bonomo}, {Desidera}, {Benatti}, {Borsa},
  {Crespi}, {Damasso}, {Lanza}, {Sozzetti}, {Lodato}, {Marzari}, {Boccato},
  {Claudi}, {Cosentino}, {Covino}, {Gratton}, {Maggio}, {Micela}, {Molinari},
  {Pagano}, {Piotto}, {Poretti}, {Smareglia}, {Affer}, {Biazzo}, {Bignamini},
  {Esposito}, {Giacobbe}, {H{\'e}brard}, {Malavolta}, {Maldonado}, {Mancini},
  {Martinez Fiorenzano}, {Masiero}, {Nascimbeni}, {Pedani}, {Rainer}, \&
  {Scandariato}}]{bonomo2017harpsMasses}
{Bonomo}, A.~S., {Desidera}, S., {Benatti}, S., {et~al.} 2017, \aap, 602, A107
 \href{https://ui.adsabs.harvard.edu/\#abs/2017A&A...602A.107B}{\adsadsurllinklabel}

\bibitem[{Bradley {et~al.}(2016)Bradley, Sipocz, Robitaille, Tollerud,
  Vin{\'\i}cius, Deil, Barbary, G{\"u}nther, Cara, Droettboom, Bostroem, Bray,
  Bratholm, Pickering, Craig, Barentsen, Pascual, adonath, Greco, Kerzendorf,
  StuartLittlefair, Ferreira, D'Eugenio, \& Weaver}]{bradley2016photutilsv0p3}
Bradley, L., Sipocz, B., Robitaille, T., {et~al.} 2016, astropy/photutils:
  v0.3, doi:10.5281/zenodo.164986
 \href{https://doi.org/10.5281/zenodo.164986}{\adsurllinklabel}

\bibitem[{Bushouse {et~al.}(2022)Bushouse, Eisenhamer, Dencheva, Davies,
  Greenfield, Morrison, Hodge, Simon, Grumm, Droettboom, Slavich, Sosey, Pauly,
  Miller, Jedrzejewski, Hack, Davis, Crawford, Law, Gordon, Regan, Cara,
  MacDonald, Bradley, Shanahan, \& Jamieson}]{bushouse2022jwstPipeline}
Bushouse, H., Eisenhamer, J., Dencheva, N., {et~al.} 2022, Github,
  doi:10.5281/zenodo.7038885
 \href{https://doi.org/10.5281/zenodo.7038885}{\adsurllinklabel}

\bibitem[{{Canipe} {et~al.}(2017){Canipe}, {Robberto}, \&
  {Hilbert}}]{canipe2014nonlinCorrection}
{Canipe}, A., {Robberto}, M., \& {Hilbert}, B. 2017, JWST-STScI-005167, SM-12
 \href{https://www.stsci.edu/files/live/sites/www/files/home/jwst/documentation/technical-documents/_documents/JWST-STScI-005167.pdf}{\adsurllinklabel}

\bibitem[{{Croll} {et~al.}(2011){Croll}, {Lafreniere}, {Albert},
  {Jayawardhana}, {Fortney}, \& {Murray}}]{croll2011wasp12}
{Croll}, B., {Lafreniere}, D., {Albert}, L., {et~al.} 2011, \aj, 141, 30
 \href{http://adsabs.harvard.edu/abs/2011AJ....141...30C}{\adsadsurllinklabel}

\bibitem[{{Dean} \&
  {Bowers}(2003)}]{dean2003diversitySelectionForPhaseRetrieval}
{Dean}, B.~H., \& {Bowers}, C.~W. 2003, Journal of the Optical Society of
  America A, 20, 1490
 \href{https://ui.adsabs.harvard.edu/abs/2003JOSAA..20.1490D}{\adsadsurllinklabel}

\bibitem[{{Deming} {et~al.}(2013){Deming}, {Wilkins}, {McCullough}, {Burrows},
  {Fortney}, {Agol}, {Dobbs-Dixon}, {Madhusudhan}, {Crouzet}, {Desert},
  {Gilliland}, {Haynes}, {Knutson}, {Line}, {Magic}, {Mandell}, {Ranjan},
  {Charbonneau}, {Clampin}, {Seager}, \& {Showman}}]{deming13}
{Deming}, D., {Wilkins}, A., {McCullough}, P., {et~al.} 2013, \apj, 774, 95
 \href{http://adsabs.harvard.edu/abs/2013ApJ...774...95D}{\adsadsurllinklabel}

\bibitem[{{Demory} {et~al.}(2022){Demory}, {Sulis}, {Meier Valdes}, {Delrez},
  {Brandeker}, {Billot}, {Fortier}, {Hoyer}, {Sousa}, {Heng}, {Lendl}, {Krenn},
  {Morris}, {Patel}, {Alibert}, {Alonso}, {Anglada}, {Barczy}, {Barrado},
  {Barros}, {Baumjohann}, {Beck}, {Beck}, {Benz}, {Bonfils}, {Broeg}, {Buder},
  {Cabrera}, {Charnoz}, {Collier Cameron}, {Cottard}, {Csizmadia}, {Davies},
  {Deleuil}, {Demangeon}, {Ehrenreich}, {Erikson}, {Fossati}, {Fridlund},
  {Gandolfi}, {Gillon}, {Gudel}, {Isaak}, {Kiss}, {Laskar}, {Lecavelier des
  Etangs}, {Lovis}, {Luntzer}, {Magrin}, {Marafatto}, {Maxted}, {Nascimbeni},
  {Olofsson}, {Ottensamer}, {Pagano}, {Palle}, {Peter}, {Piotto}, {Pollacco},
  {Queloz}, {Ragazzoni}, {Rando}, {Ratti}, {Rauer}, {Ribas}, {Santos},
  {Scandariato}, {Segransan}, {Simon}, {Smith}, {Steller}, {Szabo}, {Thomas},
  {Udry}, {Van Grootel}, \& {Walton}}]{demory2022CHEOPS55Cnce}
{Demory}, B.~O., {Sulis}, S., {Meier Valdes}, E., {et~al.} 2022, arXiv
  e-prints, arXiv:2211.03582
 \href{https://ui.adsabs.harvard.edu/abs/2022arXiv221103582D}{\adsadsurllinklabel}

\bibitem[{{Doyon} {et~al.}(2012){Doyon}, {Hutchings}, {Beaulieu}, {Albert},
  {Lafreni{\`e}re}, {Willott}, {Touahri}, {Rowlands}, {Maszkiewicz},
  {Fullerton}, {Volk}, {Martel}, {Chayer}, {Sivaramakrishnan}, {Abraham},
  {Ferrarese}, {Jayawardhana}, {Johnstone}, {Meyer}, {Pipher}, \&
  {Sawicki}}]{doyon2012NIRISSFGS}
{Doyon}, R., {Hutchings}, J.~B., {Beaulieu}, M., {et~al.} 2012, in Society of
  Photo-Optical Instrumentation Engineers (SPIE) Conference Series, Vol. 8442,
  Space Telescopes and Instrumentation 2012: Optical, Infrared, and Millimeter
  Wave, 84422R
 \href{https://ui.adsabs.harvard.edu/abs/2012SPIE.8442E..2RD}{\adsadsurllinklabel}

\bibitem[{{Espinoza} {et~al.}(2022){Espinoza}, {{\'U}beda}, {Birkmann},
  {Ferruit}, {Valenti}, {Sing}, {Rustamkulov}, {Regan}, {Kendrew}, {Sabbi},
  {Schlawin}, {Beatty}, {Albert}, {Greene}, {Nikolov}, {Karakla}, {Keyes},
  {Kumari}, {Alves de Oliveira}, {B{\"o}ker}, {Pe{\~n}a-Guerrero}, {Giardino},
  {Manjavacas}, {Proffitt}, \& {Rawle}}]{espinoza2022hatp14Spec}
{Espinoza}, N., {{\'U}beda}, L., {Birkmann}, S.~M., {et~al.} 2022, arXiv
  e-prints, arXiv:2211.01459
 \href{https://ui.adsabs.harvard.edu/abs/2022arXiv221101459E}{\adsadsurllinklabel}

\bibitem[{{Fienup}(1982)}]{fienup1982}
{Fienup}, J.~R. 1982, \ao, 21, 2758
 \href{https://ui.adsabs.harvard.edu/abs/1982ApOpt..21.2758F}{\adsadsurllinklabel}

\bibitem[{{Fienup}(1993)}]{fienup1993a}
---. 1993, \ao, 32, 1737
 \href{https://ui.adsabs.harvard.edu/abs/1993ApOpt..32.1737F}{\adsadsurllinklabel}

\bibitem[{{Fienup}(1999)}]{fienup1999}
---. 1999, Journal of the Optical Society of America A, 16, 1831
 \href{https://ui.adsabs.harvard.edu/abs/1999JOSAA..16.1831F}{\adsadsurllinklabel}

\bibitem[{{Fukui} {et~al.}(2016){Fukui}, {Narita}, {Kawashima}, {Kusakabe},
  {Onitsuka}, {Ryu}, {Ikoma}, {Yanagisawa}, \& {Izumiura}}]{fukui2016hatp14}
{Fukui}, A., {Narita}, N., {Kawashima}, Y., {et~al.} 2016, \apj, 819, 27
 \href{https://ui.adsabs.harvard.edu/abs/2016ApJ...819...27F}{\adsadsurllinklabel}

\bibitem[{{Greene} {et~al.}(2010){Greene}, {Beichman}, {Gully-Santiago},
  {Jaffe}, {Kelly}, {Krist}, {Rieke}, \& {Smith}}]{greene2010jwstNIRCam}
{Greene}, T., {Beichman}, C., {Gully-Santiago}, M., {et~al.} 2010, in Society
  of Photo-Optical Instrumentation Engineers (SPIE) Conference Series, Vol.
  7731, Space Telescopes and Instrumentation 2010: Optical, Infrared, and
  Millimeter Wave, ed. J.~{Oschmann}, Jacobus~M., M.~C. {Clampin}, \& H.~A.
  {MacEwen}, 77310C
 \href{https://ui.adsabs.harvard.edu/abs/2010SPIE.7731E..0CG}{\adsadsurllinklabel}

\bibitem[{{Greene} {et~al.}(2016){Greene}, {Line}, {Montero}, {Fortney},
  {Lustig-Yaeger}, \& {Luther}}]{greene2016jwst_trans}
{Greene}, T.~P., {Line}, M.~R., {Montero}, C., {et~al.} 2016, \apj, 817, 17
 \href{http://adsabs.harvard.edu/abs/2016ApJ...817...17G}{\adsadsurllinklabel}

\bibitem[{{Greene} {et~al.}(2017){Greene}, {Kelly}, {Stansberry}, {Leisenring},
  {Egami}, {Schlawin}, {Chu}, {Hodapp}, \& {Rieke}}]{greene2017jatisNIRCam}
{Greene}, T.~P., {Kelly}, D.~M., {Stansberry}, J., {et~al.} 2017, Journal of
  Astronomical Telescopes, Instruments, and Systems, 3, 035001
 \href{http://adsabs.harvard.edu/abs/2017JATIS...3c5001G}{\adsadsurllinklabel}

\bibitem[{{Guizar-Sicairos} {et~al.}(2008){Guizar-Sicairos}, {Thurman}, \&
  {Fienup}}]{guizar-sicairos2008}
{Guizar-Sicairos}, M., {Thurman}, S.~T., \& {Fienup}, J.~R. 2008, Optics
  Letters, 33, 156
 \href{https://ui.adsabs.harvard.edu/abs/2008OptL...33..156G}{\adsadsurllinklabel}

\bibitem[{Hunter(2007)}]{Hunter2007matplotlib}
Hunter, J.~D. 2007, Computing In Science \& Engineering, 9, 90


\bibitem[{{Ingalls} {et~al.}(2016){Ingalls}, {Krick}, {Carey}, {Stauffer},
  {Lowrance}, {Grillmair}, {Buzasi}, {Deming}, {Diamond-Lowe}, {Evans},
  {Morello}, {Stevenson}, {Wong}, {Capak}, {Glaccum}, {Laine}, {Surace}, \&
  {Storrie-Lombardi}}]{ingalls2016spitzerRepeatability}
{Ingalls}, J.~G., {Krick}, J.~E., {Carey}, S.~J., {et~al.} 2016, \aj, 152, 44
 \href{http://adsabs.harvard.edu/abs/2016AJ....152...44I}{\adsadsurllinklabel}

\bibitem[{{Jenkins} {et~al.}(2010){Jenkins}, {Caldwell}, {Chandrasekaran},
  {Twicken}, {Bryson}, {Quintana}, {Clarke}, {Li}, {Allen}, {Tenenbaum}, {Wu},
  {Klaus}, {Van Cleve}, {Dotson}, {Haas}, {Gilliland}, {Koch}, \&
  {Borucki}}]{jenkins2010initialCharKeplerLC}
{Jenkins}, J.~M., {Caldwell}, D.~A., {Chandrasekaran}, H., {et~al.} 2010,
  \apjl, 713, L120
 \href{https://ui.adsabs.harvard.edu/abs/2010ApJ...713L.120J}{\adsadsurllinklabel}

\bibitem[{{Jurling} {et~al.}(2018){Jurling}, {Bergkoetter}, \&
  {Fienup}}]{jurling2018}
{Jurling}, A.~S., {Bergkoetter}, M.~D., \& {Fienup}, J.~R. 2018, Journal of the
  Optical Society of America A, 35, 1784
 \href{https://ui.adsabs.harvard.edu/abs/2018JOSAA..35.1784J}{\adsadsurllinklabel}

\bibitem[{{Jurling} \& {Fienup}(2014)}]{jurling2014b}
{Jurling}, A.~S., \& {Fienup}, J.~R. 2014, Journal of the Optical Society of
  America A, 31, 1348
 \href{https://ui.adsabs.harvard.edu/abs/2014JOSAA..31.1348J}{\adsadsurllinklabel}

\bibitem[{{Kempton} {et~al.}(2017){Kempton}, {Bean}, \&
  {Parmentier}}]{kempton2017hazeVsCloud}
{Kempton}, E.~M.-R., {Bean}, J.~L., \& {Parmentier}, V. 2017, \apjl, 845, L20
 \href{http://adsabs.harvard.edu/abs/2017ApJ...845L..20K}{\adsadsurllinklabel}

\bibitem[{{Kendrew} {et~al.}(2015){Kendrew}, {Scheithauer}, {Bouchet},
  {Amiaux}, {Azzollini}, {Bouwman}, {Chen}, {Dubreuil}, {Fischer}, {Glasse},
  {Greene}, {Lagage}, {Lahuis}, {Ronayette}, {Wright}, \&
  {Wright}}]{kendrew2015LRSMIRI}
{Kendrew}, S., {Scheithauer}, S., {Bouchet}, P., {et~al.} 2015, \pasp, 127, 623
 \href{http://adsabs.harvard.edu/abs/2015PASP..127..623K}{\adsadsurllinklabel}

\bibitem[{{Kipping}(2013)}]{kipping2013uninfomativePriorsQuad}
{Kipping}, D.~M. 2013, \mnras, 435, 2152
 \href{https://ui.adsabs.harvard.edu/abs/2013MNRAS.435.2152K}{\adsadsurllinklabel}

\bibitem[{{Kreidberg} {et~al.}(2018){Kreidberg}, {Line}, {Parmentier},
  {Stevenson}, {Louden}, {Bonnefoy}, {Faherty}, {Henry}, {Williamson},
  {Stassun}, {Bean}, {Fortney}, {Showman}, {D{\'e}sert}, \&
  {Arcangeli}}]{kreidberg2018wasp103}
{Kreidberg}, L., {Line}, M.~R., {Parmentier}, V., {et~al.} 2018, ArXiv
  e-prints, arXiv:1805.00029
 \href{http://adsabs.harvard.edu/abs/2018arXiv180500029K}{\adsadsurllinklabel}

\bibitem[{{Kurucz}(2017)}]{kurucz2017atlas9}
{Kurucz}, R.~L. 2017, {ATLAS9: Model atmosphere program with opacity
  distribution functions}, Astrophysics Source Code Library, record
  ascl:1710.017, ascl:1710.017
 \href{https://ui.adsabs.harvard.edu/abs/2017ascl.soft10017K}{\adsadsurllinklabel}

\bibitem[{{Leisenring}(2019)}]{leisenring2020pynrc0p8dev}
{Leisenring}, J. 2019, pynrc
 \href{https://pynrc.readthedocs.io/en/latest/}{\adsurllinklabel}

\bibitem[{{Leisenring} {et~al.}(2016){Leisenring}, {Rieke}, {Misselt}, \&
  {Robberto}}]{leisenring2016persistence}
{Leisenring}, J.~M., {Rieke}, M., {Misselt}, K., \& {Robberto}, M. 2016, in
  \procspie, Vol. 9915, Society of Photo-Optical Instrumentation Engineers
  (SPIE) Conference Series, 99152N
 \href{http://adsabs.harvard.edu/abs/2016SPIE.9915E..2NL}{\adsadsurllinklabel}

\bibitem[{{Luger} {et~al.}(2019){Luger}, {Agol}, {Foreman-Mackey}, {Fleming},
  {Lustig-Yaeger}, \& {Deitrick}}]{luger2019starry}
{Luger}, R., {Agol}, E., {Foreman-Mackey}, D., {et~al.} 2019, \aj, 157, 64
 \href{https://ui.adsabs.harvard.edu/abs/2019AJ....157...64L}{\adsadsurllinklabel}

\bibitem[{{Lustig-Yaeger} {et~al.}(2019){Lustig-Yaeger}, {Meadows}, \&
  {Lincowski}}]{lustig-yaeger2019detectabilityTRAPPIST-1}
{Lustig-Yaeger}, J., {Meadows}, V.~S., \& {Lincowski}, A.~P. 2019, \aj, 158, 27
 \href{https://ui.adsabs.harvard.edu/abs/2019AJ....158...27L}{\adsadsurllinklabel}

\bibitem[{{Maxted} {et~al.}(2022){Maxted}, {Ehrenreich}, {Wilson}, {Alibert},
  {Cameron}, {Hoyer}, {Sousa}, {Olofsson}, {Bekkelien}, {Deline}, {Delrez},
  {Bonfanti}, {Borsato}, {Alonso}, {Anglada Escud{\'e}}, {Barrado}, {Barros},
  {Baumjohann}, {Beck}, {Beck}, {Benz}, {Billot}, {Biondi}, {Bonfils},
  {Brandeker}, {Broeg}, {B{\'a}rczy}, {Cabrera}, {Charnoz}, {Corral Van Damme},
  {Csizmadia}, {Davies}, {Deleuil}, {Demangeon}, {Demory}, {Erikson},
  {Flor{\'e}n}, {Fortier}, {Fossati}, {Fridlund}, {Futyan}, {Gandolfi},
  {Gillon}, {Guedel}, {Guterman}, {Heng}, {Isaak}, {Kiss}, {Laskar},
  {Lecavelier des Etangs}, {Lendl}, {Lovis}, {Magrin}, {Nascimbeni},
  {Ottensamer}, {Pagano}, {Pall{\'e}}, {Peter}, {Piotto}, {Pollacco},
  {Pozuelos}, {Queloz}, {Ragazzoni}, {Rando}, {Rauer}, {Reimers}, {Ribas},
  {Salmon}, {Santos}, {Scandariato}, {Simon}, {Smith}, {Steller}, {Swayne},
  {Szab{\'o}}, {S{\'e}gransan}, {Thomas}, {Udry}, {Van Grootel}, \&
  {Walton}}]{maxted2022pycheops}
{Maxted}, P.~F.~L., {Ehrenreich}, D., {Wilson}, T.~G., {et~al.} 2022, \mnras,
  514, 77
 \href{https://ui.adsabs.harvard.edu/abs/2022MNRAS.514...77M}{\adsadsurllinklabel}

\bibitem[{{Meier Vald{\'e}s} {et~al.}(2022){Meier Vald{\'e}s}, {Morris},
  {Wells}, {Schanche}, \& {Demory}}]{meierValdes2022TESS55Cnce}
{Meier Vald{\'e}s}, E.~A., {Morris}, B.~M., {Wells}, R.~D., {Schanche}, N., \&
  {Demory}, B.~O. 2022, \aap, 663, A95
 \href{https://ui.adsabs.harvard.edu/abs/2022A&A...663A..95M}{\adsadsurllinklabel}

\bibitem[{{Mordasini} {et~al.}(2016){Mordasini}, {van Boekel}, {Molli{\`e}re},
  {Henning}, \& {Benneke}}]{mordasini2016planetFormationSpec}
{Mordasini}, C., {van Boekel}, R., {Molli{\`e}re}, P., {Henning}, T., \&
  {Benneke}, B. 2016, \apj, 832, 41
 \href{http://adsabs.harvard.edu/abs/2016ApJ...832...41M}{\adsadsurllinklabel}

\bibitem[{{Morley} {et~al.}(2017){Morley}, {Kreidberg}, {Rustamkulov},
  {Robinson}, \& {Fortney}}]{morley2017temperateEarthSizedJWST}
{Morley}, C.~V., {Kreidberg}, L., {Rustamkulov}, Z., {Robinson}, T., \&
  {Fortney}, J.~J. 2017, \apj, 850, 121
 \href{https://ui.adsabs.harvard.edu/abs/2017ApJ...850..121M}{\adsadsurllinklabel}

\bibitem[{{Ngo} {et~al.}(2015){Ngo}, {Knutson}, {Hinkley}, {Crepp}, {Bechter},
  {Batygin}, {Howard}, {Johnson}, {Morton}, \&
  {Muirhead}}]{ngo2015friendsHotJups}
{Ngo}, H., {Knutson}, H.~A., {Hinkley}, S., {et~al.} 2015, \apj, 800, 138
 \href{https://ui.adsabs.harvard.edu/abs/2015ApJ...800..138N}{\adsadsurllinklabel}

\bibitem[{{Perrin} {et~al.}(2014){Perrin}, {Sivaramakrishnan}, {Lajoie},
  {Elliott}, {Pueyo}, {Ravindranath}, \& {Albert}}]{perrin2014webbpsf}
{Perrin}, M.~D., {Sivaramakrishnan}, A., {Lajoie}, C.-P., {et~al.} 2014, in
  \procspie, Vol. 9143, Space Telescopes and Instrumentation 2014: Optical,
  Infrared, and Millimeter Wave, 91433X
 \href{http://adsabs.harvard.edu/abs/2014SPIE.9143E..3XP}{\adsadsurllinklabel}

\bibitem[{{Perrin} {et~al.}(2016){Perrin}, {Acton}, {Lajoie}, {Knight},
  {Lallo}, {Allen}, {Baggett}, {Barker}, {Comeau}, {Coppock}, {Dean}, {Hartig},
  {Hayden}, {Jordan}, {Jurling}, {Kulp}, {Long}, {McElwain}, {Meza}, {Nelan},
  {Soummer}, {Stansberry}, {Stark}, {Telfer}, {Welsh}, {Zielinski}, \&
  {Zimmerman}}]{perrin2016preparingJWSTwavefrontSensing}
{Perrin}, M.~D., {Acton}, D.~S., {Lajoie}, C.-P., {et~al.} 2016, in Society of
  Photo-Optical Instrumentation Engineers (SPIE) Conference Series, Vol. 9904,
  Space Telescopes and Instrumentation 2016: Optical, Infrared, and Millimeter
  Wave, ed. H.~A. {MacEwen}, G.~G. {Fazio}, M.~{Lystrup}, N.~{Batalha},
  N.~{Siegler}, \& E.~C. {Tong}, 99040F
 \href{https://ui.adsabs.harvard.edu/abs/2016SPIE.9904E..0FP}{\adsadsurllinklabel}

\bibitem[{{Plazas} {et~al.}(2017){Plazas}, {Shapiro}, {Smith}, {Rhodes}, \&
  {Huff}}]{plazas2017nonlinearityAndPixelShifting}
{Plazas}, A.~A., {Shapiro}, C., {Smith}, R., {Rhodes}, J., \& {Huff}, E. 2017,
  Journal of Instrumentation, 12, C04009
 \href{https://ui.adsabs.harvard.edu/abs/2017JInst..12C4009P}{\adsadsurllinklabel}

\bibitem[{{Pont} {et~al.}(2006){Pont}, {Zucker}, \&
  {Queloz}}]{pont2006redNoiseTransits}
{Pont}, F., {Zucker}, S., \& {Queloz}, D. 2006, \mnras, 373, 231
 \href{https://ui.adsabs.harvard.edu/abs/2006MNRAS.373..231P}{\adsadsurllinklabel}

\bibitem[{{Powell} {et~al.}(2019){Powell}, {Louden}, {Kreidberg}, {Zhang},
  {Gao}, \& {Parmentier}}]{powell2019transitSignaturesHotJups}
{Powell}, D., {Louden}, T., {Kreidberg}, L., {et~al.} 2019, \apj, 887, 170
 \href{https://ui.adsabs.harvard.edu/abs/2019ApJ...887..170P}{\adsadsurllinklabel}

\bibitem[{{Rauscher} {et~al.}(2007){Rauscher}, {Fox}, {Ferruit}, {Hill},
  {Waczynski}, {Wen}, {Xia-Serafino}, {Mott}, {Alexander}, {Brambora}, {Derro},
  {Engler}, {Garrison}, {Johnson}, {Manthripragada}, {Marsh}, {Marshall},
  {Martineau}, {Shakoorzadeh}, {Wilson}, {Roher}, {Smith}, {Cabelli},
  {Garnett}, {Loose}, {Wong-Anglin}, {Zandian}, {Cheng}, {Ellis}, {Howe},
  {Jurado}, {Lee}, {Nieznanski}, {Wallis}, {York}, {Regan}, {Hall}, {Hodapp},
  {B{\"o}ker}, {De Marchi}, {Jakobsen}, \& {Strada}}]{rauscher2007detectors}
{Rauscher}, B.~J., {Fox}, O., {Ferruit}, P., {et~al.} 2007, \pasp, 119, 768
 \href{http://adsabs.harvard.edu/abs/2007PASP..119..768R}{\adsadsurllinklabel}

\bibitem[{{Rieke} {et~al.}(2005){Rieke}, {Kelly}, \&
  {Horner}}]{rieke2005nircamSPIE}
{Rieke}, M.~J., {Kelly}, D., \& {Horner}, S. 2005, in SPIE Conf Series, Vol.
  5904, Cryogenic Optical Systems and Instruments XI, ed. J.~B. {Heaney} \&
  L.~G. {Burriesci}, 1--8
 \href{http://adsabs.harvard.edu/abs/2005SPIE.5904....1R}{\adsadsurllinklabel}

\bibitem[{{Rigby} {et~al.}(2022){Rigby}, {Perrin}, {McElwain}, {Kimble},
  {Friedman}, {Lallo}, {Doyon}, {Feinberg}, {Ferruit}, {Glasse}, {Rieke},
  {Rieke}, {Wright}, {Willott}, {Colon}, {Milam}, {Neff}, {Stark}, {Valenti},
  {Abell}, {Abney}, {Abul-Huda}, {Acton}, {Adams}, {Adler}, {Aguilar}, {Ahmed},
  {Albert}, {Alberts}, {Aldridge}, {Allen}, {Altenburg}, {Alves de Oliveira},
  {Anderson}, {Anderson}, {Anderson}, {Argyriou}, {Armstrong}, {Arribas},
  {Artigau}, {Arvai}, {Atkinson}, {Bacon}, {Bair}, {Banks}, {Barrientes},
  {Barringer}, {Bartosik}, {Bast}, {Baudoz}, {Beatty}, {Bechtold}, {Beck},
  {Bergeron}, {Bergkoetter}, {Bhatawdekar}, {Birkmann}, {Blazek}, {Blome},
  {Boccaletti}, {Boeker}, {Boia}, {Bonaventura}, {Bond}, {Bosley}, {Boucarut},
  {Bourque}, {Bouwman}, {Bower}, {Bowers}, {Boyer}, {Brady}, {Braun}, {Breda},
  {Bresnahan}, {Bright}, {Britt}, {Bromenschenkel}, {Brooks}, {Brooks},
  {Brown}, {Brown}, {Brown}, {Bunker}, {Burger}, {Bushouse}, {Cale}, {Cameron},
  {Cameron}, {Canipe}, {Caplinger}, {Caputo}, {Carey}, {Carniani},
  {Carrasquilla}, {Carruthers}, {Case}, {Chance}, {Chapman}, {Charlot},
  {Charlow}, {Chayer}, {Chen}, {Cherinka}, {Chichester}, {Chilton}, {Chonis},
  {Clark}, {Clark}, {Coe}, {Coleman}, {Comber}, {Comeau}, {Connolly}, {Cooper},
  {Cooper}, {Coppock}, {Correnti}, {Cossou}, {Coulais}, {Coyle}, {Cracraft},
  {Curti}, {Cuturic}, {Davis}, {Davis}, {Dean}, {DeLisa}, {deMeester},
  {Dencheva}, {Dencheva}, {DePasquale}, {Deschenes}, {Hunor Detre}, {Diaz},
  {Dicken}, {DiFelice}, {Dillman}, {Dixon}, {Doggett}, {Donaldson}, {Douglas},
  {DuPrie}, {Dupuis}, {Durning}, {Easmin}, {Eck}, {Edeani}, {Egami},
  {Ehrenwinkler}, {Eisenhamer}, {Eisenhower}, {Elie}, {Elliott}, {Elliott},
  {Ellis}, {Engesser}, {Espinoza}, {Etienne}, {Etxaluze}, {Falini}, {Feeney},
  {Ferry}, {Filippazzo}, {Fincham}, {Fix}, {Flagey}, {Florian}, {Flynn},
  {Fontanella}, {Ford}, {Forshay}, {Fox}, {Franz}, {Fu}, {Fullerton}, {Galkin},
  {Galyer}, {Garcia Marin}, {Gardner}, {Gardner}, {Garland}, {Gasman},
  {Gaspar}, {Gaudreau}, {Gauthier}, {Geers}, {Geithner}, {Gennaro}, {Giardino},
  {Girard}, {Giuliano}, {Glassmire}, {Glauser}, {Glazer}, {Godfrey},
  {Golimowski}, {Gollnitz}, {Gong}, {Gonzaga}, {Gordon}, {Gordon},
  {Goudfrooij}, {Greene}, {Greenhouse}, {Grimaldi}, {Groebner}, {Grundy},
  {Guillard}, {Gutman}, {Ha}, {Haderlein}, {Hagedorn}, {Hainline}, {Haley},
  {Hami}, {Hamilton}, {Hammel}, {Hansen}, {Harkins}, {Harr}, {Hart}, {Hart},
  {Hartig}, {Hashimoto}, {Haskins}, {Hathaway}, {Havey}, {Hayden}, {Hecht},
  {Heller-Boyer}, {Henry}, {Hermann}, {Hernandez}, {Hesman}, {Hicks},
  {Hilbert}, {Hines}, {Hoffman}, {Holfeltz}, {Holler}, {Hoppa}, {Hott},
  {Howard}, {Hunter}, {Hunter}, {Hurst}, {Husemann}, {Hustak}, {Ilinca Ignat},
  {Irish}, {Jackson}, {Jahromi}, {Jakobsen}, {James}, {James}, {Januszewski},
  {Jenkins}, {Jirdeh}, {Johnson}, {Johnson}, {Jones}, {Jones}, {Jones},
  {Jones}, {Jordan}, {Jordan}, {Jurczyk}, {Jurling}, {Kaleida}, {Kalmanson},
  {Kammerer}, {Kang}, {Kao}, {Karakla}, {Kavanagh}, {Kelly}, {Kendrew},
  {Kennedy}, {Kenny}, {Keski-kuha}, {Keyes}, {Kidwell}, {Kinzel}, {Kirk},
  {Kirkpatrick}, {Kirshenblat}, {Klaassen}, {Knapp}, {Knight}, {Knollenberg},
  {Koehler}, {Koekemoer}, {Kovacs}, {Kulp}, {Kumari}, {Kyprianou}, {La Massa},
  {Labador}, {Labiano}, {Lagage}, {Lajoie}, {Lallo}, {Lam}, {Lamb}, {Lambros},
  {Lampenfield}, {Langston}, {Larson}, {Law}, {Lawrence}, {Lee}, {Leisenring},
  {Lepo}, {Leveille}, {Levenson}, {Levine}, {Levy}, {Lewis}, {Lewis},
  {Libralato}, {Lightsey}, {Link}, {Liu}, {Lo}, {Lockwood}, {Logue}, {Long},
  {Long}, {Loomis}, {Lopez-Caniego}, {Alvarez}, {Love-Pruitt}, {Lucy},
  {Luetzgendorf}, {Maghami}, {Maiolino}, {Major}, {Malla}, {Malumuth},
  {Manjavacas}, {Mannfolk}, {Marrione}, {Marston}, {Martel}, {Maschmann},
  {Masci}, {Masciarelli}, {Maszkiewicz}, {Mather}, {McKenzie}, {McLean},
  {McMaster}, {Melbourne}, {Mel{\'e}ndez}, {Menzel}, {Merz}, {Meyett}, {Meza},
  {Miskey}, {Misselt}, {Moller}, {Morrison}, {Morse}, {Moseley}, {Mosier},
  {Mountain}, {Mueckay}, {Mueller}, {Mullally}, {Murphy}, {Murray}, {Murray},
  {Muzerolle}, {Mycroft}, {Myers}, {Myrick}, {Nanavati}, {Nance}, {Nayak},
  {Naylor}, {Nelan}, {Nickson}, {Nielson}, {Nieto-Santisteban}, {Nikolov},
  {Noriega-Crespo}, {O'Shaughnessy}, {O'Sullivan}, {Ochs}, {Ogle}, {Oleszczuk},
  {Olmsted}, {Osborne}, {Ottens}, {Owens}, {Pacifici}, {Pagan}, {Page},
  {Parrish}, {Patapis}, {Pauly}, {Pavlovsky}, {Pedder}, {Peek},
  {Pena-Guerrero}, {Pennanen}, {Perez}, {Perna}, {Perriello}, {Phillips},
  {Pietraszkiewicz}, {Pinaud}, {Pirzkal}, {Pitman}, {Piwowar}, {Platais},
  {Player}, {Plesha}, {Pollizi}, {Polster}, {Pontoppidan}, {Porterfield},
  {Proffitt}, {Pueyo}, {Pulliam}, {Quirt}, {Quispe Neira}, {Ramos Alarcon},
  {Ramsay}, {Rapp}, {Rapp}, {Rauscher}, {Ravindranath}, {Rawle}, {Regan},
  {Reichard}, {Reis}, {Ressler}, {Rest}, {Reynolds}, {Rhue}, {Richon},
  {Rickman}, {Ridgaway}, {Ritchie}, {Rix}, {Robberto}, {Robinson}, {Robinson},
  {Robinson}, {Rock}, {Rodriguez}, {Rodriguez Del Pino}, {Roellig}, {Rohrbach},
  {Roman}, {Romelfanger}, {Rose}, {Roteliuk}, {Roth}, {Rothwell}, {Rowlands},
  {Roy}, {Royer}, {Royle}, {Rui}, {Rumler}, {Runnels}, {Russ}, {Rustamkulov},
  {Ryden}, {Ryer}, {Sabata}, {Sabatke}, {Sabbi}, {Samuelson}, {Sappington},
  {Sargent}, {Sauer}, {Scheithauer}, {Schlawin}, {Schlitz}, {Schmitz},
  {Schneider}, {Schreiber}, {Schulze}, {Schwab}, {Scott}, {Sembach},
  {Shaughnessy}, {Shaw}, {Shawger}, {Shay}, {Sheehan}, {Shen}, {Sherman},
  {Shiao}, {Shih}, {Shivaei}, {Sienkiewicz}, {Sing}, {Sirianni},
  {Sivaramakrishnan}, {Skipper}, {Sloan}, {Slocum}, {Slowinski}, {Smith},
  {Smith}, {Smith}, {Smith}, {Snyder}, {Soh}, {Sohn}, {Soto}, {Spencer},
  {Stallcup}, {Stansberry}, {Starr}, {Starr}, {Stewart}, {Stiavelli},
  {Straughn}, {Strickland}, {Stys}, {Summers}, {Sun}, {Sunnquist}, {Swade},
  {Swam}, {Swaters}, {Swoish}, {Taylor}, {Taylor}, {Te Plate}, {Tea}, {Teague},
  {Telfer}, {Temim}, {Thatte}, {Thompson}, {Thompson}, {Thomson}, {Tikkanen},
  {Tippet}, {Todd}, {Toolan}, {Tran}, {Trejo}, {Truong}, {Tsukamoto},
  {Tustain}, {Tyra}, {Ubeda}, {Underwood}, {Uzzo}, {Van Campen}, {Vandal},
  {Vandenbussche}, {Vila}, {Volk}, {Wahlgren}, {Waldman}, {Walker}, {Wander},
  {Warfield}, {Warner}, {Wasiak}, {Watkins}, {Weilert}, {Weiser}, {Weiss},
  {Weissman}, {Welty}, {West}, {Wheate}, {Wheatley}, {Wheeler}, {White},
  {Whiteaker}, {Whitehouse}, {Whiteleather}, {Whitman}, {Williams}, {Willmer},
  {Willoughby}, {Wilson}, {Wirth}, {Wislowski}, {Wolf}, {Wolfe}, {Wolff},
  {Workman}, {Wright}, {Wu}, {Wu}, {Wymer}, {Yates}, {Yates}, {Yeager},
  {Yerger}, {Yoon}, {Young}, {Yu}, {Zak}, {Zeidler}, {Zhou}, {Zielinski},
  {Zincke}, \& {Zonak}}]{rigby2022jwstPerformance}
{Rigby}, J., {Perrin}, M., {McElwain}, M., {et~al.} 2022, arXiv e-prints,
  arXiv:2207.05632
 \href{https://ui.adsabs.harvard.edu/abs/2022arXiv220705632R}{\adsadsurllinklabel}

\bibitem[{Sahlmann {et~al.}(2019)Sahlmann, Osborne, Cox, charlesrp, Law,
  Perrin, marthaboyer, \& Hunkeler}]{sahlmann2019pysiaf}
Sahlmann, J., Osborne, S., Cox, C., {et~al.} 2019, {spacetelescope/pysiaf:
  Introduce check against online version}, doi:10.5281/zenodo.3516964


\bibitem[{{Schlawin} {et~al.}(2018){Schlawin}, {Greene}, {Line}, {Fortney}, \&
  {Rieke}}]{schlawin2018JWSTforecasts}
{Schlawin}, E., {Greene}, T.~P., {Line}, M., {Fortney}, J.~J., \& {Rieke}, M.
  2018, \aj, 156, 40
 \href{http://adsabs.harvard.edu/abs/2018AJ....156...40S}{\adsadsurllinklabel}

\bibitem[{{Schlawin} {et~al.}(2020){Schlawin}, {Leisenring}, {Misselt},
  {Greene}, {McElwain}, {Beatty}, \& {Rieke}}]{schlawin2020jwstNoiseFloorI}
{Schlawin}, E., {Leisenring}, J., {Misselt}, K., {et~al.} 2020, \aj, 160, 231
 \href{https://ui.adsabs.harvard.edu/abs/2020AJ....160..231S}{\adsadsurllinklabel}

\bibitem[{{Schlawin} {et~al.}(2017){Schlawin}, {Rieke}, {Leisenring}, {Walker},
  {Fraine}, {Kelly}, {Misselt}, {Greene}, {Line}, {Lewis}, \&
  {Stansberry}}]{schlawin2017dhs}
{Schlawin}, E., {Rieke}, M., {Leisenring}, J., {et~al.} 2017, \pasp, 129,
  015001
 \href{http://adsabs.harvard.edu/abs/2017PASP..129a5001S}{\adsadsurllinklabel}

\bibitem[{{Schlawin} {et~al.}(2021){Schlawin}, {Leisenring}, {McElwain},
  {Misselt}, {Don}, {Greene}, {Beatty}, {Nikolov}, {Kelly}, \&
  {Rieke}}]{schlawin2020jwstNoiseFloor2}
{Schlawin}, E., {Leisenring}, J., {McElwain}, M.~W., {et~al.} 2021, \aj, 161,
  115
 \href{https://ui.adsabs.harvard.edu/abs/2021AJ....161..115S}{\adsadsurllinklabel}

\bibitem[{{Smith} {et~al.}(2008){Smith}, {Zavodny}, {Rahmer}, \&
  {Bonati}}]{smith2008imgPersistence}
{Smith}, R.~M., {Zavodny}, M., {Rahmer}, G., \& {Bonati}, M. 2008, in High
  Energy, Optical, and Infrared Detectors for Astronomy III, Vol. 7021, 70210J
 \href{https://ui.adsabs.harvard.edu/#abs/2008SPIE.7021E..0JS}{\adsadsurllinklabel}

\bibitem[{{Southworth}(2012)}]{southworth2012homogenousStudies38Planets}
{Southworth}, J. 2012, \mnras, 426, 1291
 \href{https://ui.adsabs.harvard.edu/abs/2012MNRAS.426.1291S}{\adsadsurllinklabel}

\bibitem[{{Stassun} {et~al.}(2017){Stassun}, {Collins}, \&
  {Gaudi}}]{stassun2017gaiaRadiiMasses}
{Stassun}, K.~G., {Collins}, K.~A., \& {Gaudi}, B.~S. 2017, \aj, 153, 136
 \href{http://adsabs.harvard.edu/abs/2017AJ....153..136S}{\adsadsurllinklabel}

\bibitem[{{STScI}(2016)}]{jdox2016}
{STScI}. 2016, {JWST User Documentation (JDox)}, JWST User Documentation
  Website
 \href{https://ui.adsabs.harvard.edu/abs/2016jdox.rept......}{\adsadsurllinklabel}

\bibitem[{{Thurman} \& {Fienup}(2009)}]{thurman2009a}
{Thurman}, S.~T., \& {Fienup}, J.~R. 2009, Journal of the Optical Society of
  America A, 26, 1008
 \href{https://ui.adsabs.harvard.edu/abs/2009JOSAA..26.1008T}{\adsadsurllinklabel}

\bibitem[{{Torres} {et~al.}(2010){Torres}, {Bakos}, {Hartman}, {Kov{\'a}cs},
  {Noyes}, {Latham}, {Fischer}, {Johnson}, {Marcy}, {Howard}, {Sasselov},
  {Kipping}, {Sip{\H{o}}cz}, {Stefanik}, {Esquerdo}, {Everett},
  {L{\'a}z{\'a}r}, {Papp}, \& {S{\'a}ri}}]{torres2010hatp14}
{Torres}, G., {Bakos}, G.~{\'A}., {Hartman}, J., {et~al.} 2010, \apj, 715, 458
 \href{https://ui.adsabs.harvard.edu/abs/2010ApJ...715..458T}{\adsadsurllinklabel}

\bibitem[{{van der Walt} {et~al.}(2011){van der Walt}, {Colbert}, \&
  {Varoquaux}}]{vanderWalt2011numpy}
{van der Walt}, S., {Colbert}, S.~C., \& {Varoquaux}, G. 2011, Computing in
  Science and Engineering, 13, 22
 \href{https://ui.adsabs.harvard.edu/abs/2011CSE....13b..22V}{\adsadsurllinklabel}

\bibitem[{{Virtanen} {et~al.}(2020){Virtanen}, {Gommers}, {Oliphant},
  {Haberland}, {Reddy}, {Cournapeau}, {Burovski}, {Peterson}, {Weckesser},
  {Bright}, {van der Walt}, {Brett}, {Wilson}, {Jarrod Millman}, {Mayorov},
  {Nelson}, {Jones}, {Kern}, {Larson}, {Carey}, {Polat}, {Feng}, {Moore}, {Vand
  erPlas}, {Laxalde}, {Perktold}, {Cimrman}, {Henriksen}, {Quintero}, {Harris},
  {Archibald}, {Ribeiro}, {Pedregosa}, {van Mulbregt}, \&
  {Contributors}}]{virtanen2020scipy}
{Virtanen}, P., {Gommers}, R., {Oliphant}, T.~E., {et~al.} 2020, Nature
  Methods, 17, 261
 \href{https://rdcu.be/b08Wh}{\adsurllinklabel}

\bibitem[{{Wakeford} {et~al.}(2019){Wakeford}, {Wilson}, {Stevenson}, \&
  {Lewis}}]{wakeford2019atmosphericForecastMuted}
{Wakeford}, H.~R., {Wilson}, T.~J., {Stevenson}, K.~B., \& {Lewis}, N.~K. 2019,
  Research Notes of the American Astronomical Society, 3, 7
 \href{https://ui.adsabs.harvard.edu/abs/2019RNAAS...3....7W}{\adsadsurllinklabel}

\bibitem[{{Zhou} {et~al.}(2017){Zhou}, {Apai}, {Lew}, \&
  {Schneider}}]{zhou2017chargeTrap}
{Zhou}, Y., {Apai}, D., {Lew}, B.~W.~P., \& {Schneider}, G. 2017, \aj, 153, 243
 \href{http://adsabs.harvard.edu/abs/2017AJ....153..243Z}{\adsadsurllinklabel}

\end{thebibliography}



\end{document}